\documentclass[10pt,a4paper]{iopart}
\usepackage{iopams}  
\usepackage{graphicx,psfrag,bbm,latexsym,color,dcolumn,bm,dsfont,bbm,color,mathrsfs,bbold,latexsym,amsfonts,amssymb}

\newcommand{\ua}{\uparrow}
\newcommand{\da}{\downarrow}
\newcommand{\media}[1]{\left\langle #1 \right\rangle_N}

\newcommand{\mediac}[1]{\left\langle #1 \right\rangle_{N}^{(c)}}
\newcommand{\E}{\mbox{$\mathsf E$}}

\def\mod{ \mathop{\rm mod} }
\def\Re{ \mathop{\rm Re} }
\def\Im{ \mathop{\rm Im} }
\def\Arg{ \mathop{\rm Arg} }

\begin{document}
\title[Ground state of lattice quantum systems: cumulant expansion]{Analytical probabilistic approach to the ground state of lattice 
quantum systems: exact results in terms of a cumulant expansion}

\author{Massimo Ostilli$^{1,2}$ and Carlo Presilla$^{1,2,3}$}

\address{$^1$\ Dipartimento di Fisica, Universit\`a di Roma ``La Sapienza'', Piazzale A. Moro 2, Roma 00185, Italy}
\address{$^2$\ Center for Statistical Mechanics and Complexity, Istituto Nazionale per la Fisica della Materia, Unit\`a di Roma 1, Roma 00185, Italy}
\address{$^3$\ Istituto Nazionale di Fisica Nucleare, Sezione di Roma 1, 
Roma 00185, Italy}

\date{\today}

\begin{abstract}
We present a large deviation analysis of a recently proposed probabilistic
approach to the study of the ground-state properties of lattice quantum
systems.
The ground-state energy, 
as well as the correlation functions in the ground state, 
are exactly determined as a series expansion in the cumulants 
of the multiplicities of the potential and hopping energies 
assumed by the system during its long-time evolution.
Once these cumulants are known, even at a finite order, 
our approach provides the ground state analytically as a function of 
the Hamiltonian parameters.
A scenario of possible applications of this analyticity property 
is discussed. 
\end{abstract}

\maketitle


\section{Introduction}
The real- or imaginary-time dynamics 
of systems described by a finite Hamiltonian matrix, representing
bosonic or fermionic degrees of freedom, admits an exact
probabilistic representation in terms of a proper collection 
of independent Poisson processes~\cite{DAJLS,DJS,PRESILLA}.
For a lattice system, the Poisson processes are associated 
to the links of the lattice and the probabilistic representation 
leads to an optimal algorithm~\cite{PRESILLA} which coincides with 
the Green Function Quantum Monte Carlo method 
in the limit when the latter becomes exact~\cite{SORELLACAPRIOTTI}.

Recently, in \cite{OP1} we have exploited the above 
probabilistic representation to derive analytical estimations for the matrix
elements of the evolution operator in the long-time limit.
In this way, approximations for the ground-state energy, 
as well as for the expectation of a generic operator in the ground state 
of a lattice system without sign problem, 
have been obtained as the solution of a simple scalar equation.
The results presented in \cite{OP1} were based on the application
of a central limit theorem to the rescaled multiplicities 
of the values assumed by the potential and hopping energies in the 
configurations dynamically visited by the system (see also Ref.~\cite{OP2} 
for details on the corresponding probability density).

The approximated scalar equation for the ground-state energy obtained by using 
the central limit theorem contains only the first two statistical moments
of the multiplicities above mentioned and, as anticipated in~\cite{OP1}, 
suggests that it is a second order truncation of an exact cumulant expansion.
In this paper, after reviewing in detail the approach developed in \cite{OP1}, 
we perform a large deviation analysis of the relevant 
probability density and obtain this exact cumulant expansion.
In principle, if all the cumulants are known, 
we provide a scalar equation whose straightforward solution is 
the exact ground-state energy of the system.
In practice, we measure the cumulants by a statistical sampling and  
only a finite number of them is available. 
The corresponding truncated equation provides ground-state energies whose
level of approximation depends on the values of the Hamiltonian parameters
and on the size of the system.
However, the convergence as a function of the number of cumulants is
rather fast and, as shown in some example cases, 
the use of the first few cumulants provides results indistinguishable
from those obtained by exact numerical simulations.

The spirit of our approach is different from that of Monte Carlo simulations.
In the latter case, the accumulated statistical data provide information 
on the specific system under investigation and new simulations must be 
performed each time the value of a single parameter of the model is changed. 
On the other hand, our approach is semi-analytical in the sense that
we have an exact expression for the ground-state energy, in which some
statistical data, namely the cumulants of the multiplicities above mentioned,
must be provided as an input. 
However, these multiplicities, and so the corresponding cumulants,
do not depend on the values of the parameters of the model but only on the
structure of the Hamiltonian operator
(form of the hopping and interaction terms, size of the system, and so on).
For instance, in the case of $N$ spin-up and  $N$ spin-down particles 
with on-site interaction of strength $\gamma$ 
the multiplicities depend on the spectrum of the possible on-site pairs,
$0,1,2, \dots ,N$, not on $\gamma$ that represents, 
therefore, a parameter with respect to which our results are analytical.  
The analyticity with respect to the parameters of the Hamiltonian  
allows the determination of the expectation of a generic operator 
in the ground state of the system via the Hellman-Feynman theorem and
suggests other potential applications that we will discuss at the end.

The paper is organized as follows. 
In Section \ref{representation} we briefly review the probabilistic 
representation of quantum dynamics for generic lattice systems.
In this way, the ground-state energy is obtained as the expectation
of a suitable stochastic functional.
In Section \ref{outline} we summarize our main result, namely
an exact formula for the ground-state energy, and provide an outline 
of its proof. 
In the four Subsections of the core Section \ref{pelt},
we develop in full detail this proof via the analytical determination,
in the limit of an infinitely long time, 
of the expectation of the stochastic functional providing the ground-state 
energy.
In Section \ref{decomposition}, 
we decompose the expectation in a series of canonical 
averages of weights, which are calculated in Section \ref{evaluateweights}. 
The canonical averages are evaluated
via a cumulant expansion theorem in Section \ref{largedeviation} 
and are finally resummed in Section \ref{resumming}, 
where the exact scalar equation for the ground-state energy
is presented.
In Section \ref{nresults} we discuss the numerical evaluation of the
cumulants which appear in our formula for the ground-state energy. 
In the same Section, we also study some example cases and compare 
the results from our formula with those from exact numerical calculations.
Finally, general features of our approach and future applications
are summarized and discussed in
Section \ref{conclusions}.

Up to Section \ref{evaluateweights}, we carry on the development
both for hard-core bosons and fermions, whereas successively we limit
the method to hard-core bosons postponing the fermion case 
to a later work.

\section{Exact probabilistic representation of lattice dynamics}
\label{representation}
In this Section we review the exact probabilistic representation 
of the imaginary- or real-time dynamics of a system of bosons or 
fermions on a lattice, 
see Ref.~\cite{PRESILLA} for a detailed description.
This representation is at the basis of our approach 
to study the ground-state properties of the system in a 
semi-analytical way~\cite{OP1}.
We concentrate on a simple exclusion dynamics,
in which multiple occupancies of the lattice sites are forbidden, 
\textit{i.e.} fermions or hard-core bosons are considered. 
 
Let $\Lambda$ be any finite lattice with cardinality $|\Lambda|$ 
and $1,2,\ldots,|\Lambda|$ some total ordering of the lattice sites.
The dynamics of a system of hard-core bosons or fermions on this
lattice is effectively 
described in terms of commuting or anticommuting destruction operators 
$\hat{c}^{}_{i \sigma}$ with the property 
$( \hat{c}_{i\sigma} )^2=0$.
Here, $i=1,2,\ldots,|\Lambda|$ is the site index and 
$\sigma=\ua\da$ the spin, or an arbitrary internal index. 
The system can be represented in terms of Fock states, $\bm{n}\in \mathbb{F}$,
$\bm{n}= (n_{1 \ua},n_{1 \da}, \ldots, 
n_{|\Lambda| \ua}, n_{|\Lambda| \da})$,
where  
$n_{i \sigma}$ is the lattice occupation number of the site $i$ 
with spin $\sigma$ taking the values 0 or 1.
The number of Fock states is 
\begin{eqnarray}
\left| \mathbb{F} \right| = 
{|\Lambda| \choose N_\ua} {|\Lambda| \choose N_\da},
\end{eqnarray}
where $N_{\sigma}$ is the number of particles with spin $\sigma$.

We assume the system to be described by the Hamiltonian
\begin{eqnarray}
\label{Hubbard}
\hat{H} &=& \hat{K} + \hat{V},
\end{eqnarray}
where $\hat{K}$ and $\hat{V}$ are the kinetic and potential operators, 
respectively. 
The potential operator $\hat{V}$ is arbitrary, 
\textit{e.g.} for the Hubbard model
$\hat{V} = \sum_{i \in \Lambda} \gamma_i 
\hat{c}^\dag_{i\ua} \hat{c}^{}_{i\ua} \hat{c}^\dag_{i\da} \hat{c}^{}_{i\da}$.
For the kinetic operator we assume the quadratic form
\begin{eqnarray}
\label{K}
\hat{K} &=& -
\sum_{i \neq j \in \Lambda} 
\sum_{\sigma=\ua\da} \eta_{ij\sigma} e^{i \theta_{ij\sigma}}
\hat{c}^\dag_{i\sigma} \hat{c}^{}_{j\sigma}, 
\end{eqnarray}
with $\eta_{ij\sigma}, \theta_{ij\sigma} \in \mathbb{R}$
and $\eta_{ij\sigma} \geq 0$.
The case $\theta_{ij\sigma}\neq 0$ is obtained in the presence 
of a magnetic field. 
In principle, our approach can be extended to more general kinetic operators.
The essential feature to be noted is that in the Fock
representation $\hat{V}$ is diagonal whereas $\hat{K}$ is off-diagonal.

In order to study the ground-state properties of the Hamiltonian $\hat{H}$, 
it is sufficient to evaluate the long-time behavior of 
$\sum_{\bm{n}} \langle \bm{n} |e^{- \hat{H}t} | \bm{n}_0 \rangle$.
In fact, the ground-state energy is given by
\begin{eqnarray}
\label{E0q} 
E_0 = \lim_{t \to \infty}
- \partial_t \log \sum_{\bm{n}} \langle \bm{n}|e^{-\hat{H}t} | \bm{n}_0\rangle.
\end{eqnarray}
The quantum expectation of a generic operator $\hat{O}$ 
in the ground state of $\hat{H}$ can be obtained by evaluating the 
ground-state energy $E_0(\xi)$ of the modified Hamiltonian 
$\hat{H}+\xi \hat{O}$ and using the Hellman-Feynman identity
\begin{eqnarray}
\label{HF}
\frac{\langle E_0(\xi) 
|\partial_{\xi} \hat{H}(\xi)| 
E_0(\xi)\rangle}
{\langle E_0(\xi) |E_0(\xi)\rangle}=
\partial_{\xi} E_{0}(\xi). 
\end{eqnarray}

Let $\Gamma_{\sigma}$ be the set of system links with spin $\sigma$, 
\textit{i.e.} the pairs $(i,j)$ with $i\neq j$ and  $i,j\in\Lambda$ 
such that $\eta_{ij\sigma}\neq 0$.
To each link $(i,j)$ with spin $\sigma$ we associate the value 
\begin{eqnarray}
\label{lambda0}
\lambda_{ij \sigma}(\bm{n}) = 
\langle \bm{n} \oplus \bm{1}_{i\sigma} \oplus \bm{1}_{j\sigma} |
e^{i \theta_{ij\sigma}} \hat{c}^\dag_{i\sigma} \hat{c}^{}_{j\sigma} 
+ 
e^{i \theta_{ji\sigma}} \hat{c}^\dag_{j\sigma} \hat{c}^{}_{i\sigma} 
|\bm{n}\rangle, 
\end{eqnarray}
where $\bm{1}_{i\sigma}=(0,\ldots,0,1_{i\sigma},0,\ldots,0)$
and $n \oplus n' = (n+n') \mod 2$.
We may have $|\lambda_{ij\sigma}|=0,1$ only. 
A link $(i,j)$ with spin $\sigma$ is called active if
$\lambda_{ij\sigma}\neq 0$.
For $\theta_{ij\sigma} \equiv 0$,
in the case of hard-core bosons an active link
can take only the positive value $\lambda_{ij\sigma}=1$, 
whereas in the case of fermions an active link is either $1$ or $-1$,
depending whether the number of particles present between the sites 
$i$ and $j$, $\sum_{k=i+1}^{j-1}n_{k\sigma}$, 
is even or odd.

Let $\left\{ N^t_{ij\sigma} \right\}$ be a family of 
$|\Gamma|=|\Gamma_\ua \cup \Gamma_\da|$ independent Poisson process 
with jump rate $\rho$. 
These processes are in a one-to-one correspondence 
with the links of the lattice.
At each jump of a Poisson process relating sites $i$ and $j$ with 
spin $\sigma$ and taking place at a given configuration $\bm{n}$,
a particle of spin $\sigma$ moves from site $i$ to site $j$
or vice versa if $|\lambda_{ij\sigma} (\bm{n})|= 1$, 
whereas the lattice configuration $\bm{n}$ remains 
unchanged if $\lambda_{ij\sigma} (\bm{n}) = 0$.
By arranging the jumps according to the times $s_{k}$, $k=1,\dots, N_{t}$, 
at which they take place in the interval $[0,t)$,
we define a trajectory as the sequence of configurations
$\bm{n}_{1}, \bm{n}_2, \dots, \bm{n}_{N_{t}}$ 
generated from a chosen initial configuration $\bm{n}_0$
by exchanging, at each jump of the Poisson process $N^t_{ij\sigma}$,
the occupations of the sites $n_{i\sigma}$ and $n_{j\sigma}$.
The number of jumps $N_t$ is, of course, a random integer associated to
each trajectory.
Let $(i_k,j_k)$ with spin $\sigma_k$ be the link jumping at the time $s_k$.
By putting for brevity
\begin{eqnarray}
\label{lambda}
\lambda_k=\lambda_{i_kj_k\sigma_k}(\bm{n}_{k-1}),
\end{eqnarray} 
\begin{eqnarray}
\label{eta}
\eta_k=\eta_{i_kj_k\sigma_k},
\end{eqnarray} 
we define with $\lambda_{1},\lambda_{2}, \dots, \lambda_{N_{t}}$, 
and $\eta_{1},\eta_{2}, \dots, \eta_{N_{t}}$,
the sequences of the corresponding link values and hopping parameters.
To each trajectory, we also associate the sequences,
$A_{0},A_{1}, \dots, A_{N_{t}}$ and $V_{0},V_{1} \dots, V_{N_{t}}$, 
representing the number of active links and the potential energy 
of the visited configurations
\begin{eqnarray}
\label{A}
A_k = 
\sum_{\sigma=\ua\da} \sum_{(i,j) \in \Gamma_\sigma} 
|\lambda_{ij\sigma}(\bm{n}_k)|
\end{eqnarray} 
\begin{eqnarray}
\label{V}
V_k = \langle \bm{n}_k|\hat{V}| \bm{n}_k \rangle.
\end{eqnarray}
For later use, we also define 
\begin{eqnarray}
\label{T}
T_k = A_k \eta_{k+1} /\epsilon ,
\end{eqnarray}
where $\epsilon$ is an arbitrary reference energy.

At any finite time $t$, 
the following exact probabilistic representation holds~\cite{PRESILLA}
\begin{eqnarray}
\label{TheFormulaa}
\langle \bm{n}|e^{-Ht} | \bm{n}_0\rangle &=&  
\E  \left( \delta_{ \bm{n} , \bm{n}_{N_t}} 
{\cal M}^t_{\bm{n}_0} \right), 
\end{eqnarray}
where the stochastic functional ${\cal M}^t_{\bm{n}_0}$ is defined as
\begin{eqnarray}
\label{FORMULA D}
{\cal M}^t_{\bm{n}_0} =
e^{|\Gamma| \rho t}
\left( 
\prod_{k=1}^{N_{t}} \frac{\eta_k}{\rho} \lambda_k
e^{-V_{k-1}(s_{k}-s_{k-1})} 
\right)  
e^{-V_{N_{t}}(t-s_{N_{t}})}
\end{eqnarray}
if $N_t > 0$ and 
${\cal M}^t_{\bm{n}_0} = e^{|\Gamma| \rho t} e^{-V_0t}$ if $N_t=0$
and the symbol $\E(\cdot)$ means expectation over the $|\Gamma|$ 
independent Poisson processes.
In Eq.~(\ref{FORMULA D}), we put $s_0=0$. 
According to this representation, we have 
\begin{eqnarray}
\label{TheFormulab}
\sum_{\bm{n}} \langle \bm{n}|e^{-Ht} | \bm{n}_0\rangle =  
\E  \left( {\cal M}^t_{\bm{n}_0} \right),  
\end{eqnarray}
so that we can evaluate the ground-state energy as
\begin{eqnarray}
\label{E0} 
E_0 
= \lim_{t \to \infty} -\partial_t \log 
\E \left( \mathcal{M}^t_{\bm{n}_0} \right).
\end{eqnarray}

In the following, we will be able to reduce the study of the expectation 
of the stochastic functional (\ref{FORMULA D}) to the study of simpler
canonical averages over the stochastic trajectories 
$\bm{n}_{0},\bm{n}_{1}, \bm{n}_2, \dots$ 
in which $\bm{n}_k \neq \bm{n}_{k-1}$ for any $k>1$,
\textit{i.e.} trajectories in which only jumps corresponding to active links 
are considered,
and the jumping times are disregarded (integrated out). 
These sequences of configurations 
$\bm{n}_{0},\bm{n}_{1}, \bm{n}_2, \dots$ 
constitute a homogeneous Markov chain in a finite state space, $\mathbb{F}$, 
with transition matrix $\bm{P}$
defined by  
\begin{eqnarray}\fl
\label{MARKOV}
P_{\bm{n},\bm{n}'} = \left\{
\begin{array}{ll}
A(\bm{n})^{-1} &  
\mbox{
\begin{minipage}[t]{3in}
if $\exists~\sigma \in \{\uparrow,\downarrow\}$  
and $(i,j) \in \Gamma_\sigma$:
$\bm{n}'=\bm{n} \oplus \bm{1}_{i\sigma} \oplus \bm{1}_{j\sigma}$ 
and $\lambda_{ij\sigma}(\bm{n}) \neq 0$
\end{minipage}}
\\
0 & \mbox{
\begin{minipage}[t]{3in}
otherwise,
\end{minipage}}
\end{array}
\right.
\end{eqnarray} 
where $A(\bm{n})=\sum_{\sigma=\ua\da} \sum_{(i,j) \in \Gamma_\sigma} 
|\lambda_{ij\sigma}(\bm{n})|
$ is the number of active links of the configuration $\bm{n}$.
Note that $\sum_{\bm{n}'} P_{\bm{n},\bm{n}'} =1$ according to the
fact that $P_{\bm{n},\bm{n}'}$ is the probability for the 
transition $\bm{n} \to \bm{n}'$.
Finally, the canonical averages mentioned above are introduced 
for trajectories with a finite number of jumps in the following way. 
Given an initial state $\bm{n}_0$ and an application 
$f:\mathbb{F}^{N+1} \rightarrow \mathbb{C}$,  
function of $\bm{n}_0$ and of the consecutive $N$ configurations  
$\bm{n}_1,\bm{n}_2 \ldots, \bm{n}_N$,
we will indicate with $\media{f}$, called canonical average, 
the average of $f$ sampled with respect to the measure induced by 
the transition probability (\ref{MARKOV}) iterated $N$ times. 
Similarly, we can consider canonical averages in
which the initial state is not a single Fock state, $\bm{n}_0$, 
but an ensemble with distribution $\bm{\pi}_0$.
An ensemble of particular interest is the invariant measure $\bm{\pi}$,
defined as the left eigenvector of the transition matrix, 
$\bm{\pi}^{\mathrm{T}} \bm{P} = \bm{\pi}^{\mathrm{T}}$. 
It is simple to verify that 
\begin{eqnarray}
\label{PI} 
\pi(\bm{n})=\frac{A(\bm{n})}{\sum_{\bm{n}'}A(\bm{n}')}.
\end{eqnarray}

\section{Main result and outline of the proof}
\label{outline}

Evaluating the expectation 
$\E  \left( \mathcal{M}^t_{\bm{n}_0} \right)$ 
over the detailed realizations of the stochastic processes specified above
can be done numerically by a Monte Carlo method~\cite{PRESILLA}.
The efficiency of the corresponding numerical algorithm is 
discussed in detail in Ref.~\cite{OP3}.
In this paper, we are interested to obtain
analytical expressions of $\E\left( \mathcal{M}^t_{\bm{n}_0}\right)$,
which are asymptotically exact in the limit $t \to \infty$, 
and provide an exact formula for the ground-state energy 
by using Eq. (\ref{E0}).

Let us call $\mathscr{L}$, $\mathscr{V}$ and $\mathscr{T}$ 
the sets of all the possible different values $\lambda$, $V$ and $T$ assumed 
by Eqs.~(\ref{lambda}), (\ref{V}) and (\ref{T}), respectively,
along a trajectory $\bm{n}_{0},\bm{n}_{1}, \bm{n}_2, \dots $ 
formed by infinitely many jumps.
Let $m_\mathscr{L}$, $m_\mathscr{V}$ and $m_\mathscr{T}$ be
the corresponding cardinalities.
Since any configuration can be obtained from any other one by a finite 
number of jumps, 
\textit{i.e.} the Markov chain of the trajectories is irreducible, 
the elements in the sets $\mathscr{L}$, $\mathscr{V}$ and $\mathscr{T}$
do not depend on the initial configuration $\bm{n}_0$.
Moreover, the value of their elements and, in particular, their number 
depend only on the structure of the Hamiltonian operator, 
not on the values of the Hamiltonian parameters.

As we shall show, a crucial point is that, if we consider
the conditional expectation in which the number of jumps $N$ is
fixed, $\E \left( {\cal M}^{t}_{\bm{n}_0}|N_{t}=N \right)$,
and integrate over all the
possible jump times, what matters in determining the value of
$\E \left( {\cal M}^{t}_{\bm{n}_0}|N_{t}=N \right)$,
is not the detailed sequences of the visited configurations but only
the multiplicities, or numbers of occurrences,  
$N_{\lambda}$, $N_V$ and $N_T$ of 
the variables $\lambda$, $V$ and $T$, respectively.
Explicitly, for a trajectory with $N$ jumps these multiplicities are 
defined as
\numparts
\begin{eqnarray}
N_{\lambda} = \sum_{k=0}^{N-1} \delta^{}_{\lambda_{k},\lambda}, 
\label{MULTIPLICITY-L} \\
N_{V} = \sum_{k=0}^N \delta^{}_{V_{k},V}, 
\label{MULTIPLICITY-V} \\
N_{T} = \sum_{k=0}^{N-1} \delta^{}_{T_{k},T}, 
\label{MULTIPLICITY-T}
\end{eqnarray}
\endnumparts
with $\lambda_k$, $V_k$ and $T_k$ given by Eqs.~(\ref{lambda}), (\ref{V})
and (\ref{T}).
The expectation $\E \left( {\cal M}^{t}_{\bm{n}_0}|N_{t}=N \right)$ is 
thus reduced to an average over the variables $N_{\lambda}$, $N_V$ and $N_T$.
In other words, after the stochastic times have been integrated out, 
the knowledge of all the canonical moments 
\begin{eqnarray}
\media{N_{\alpha_1} \dots N_{\alpha_k}}, 
\end{eqnarray}
where 
$\alpha_{i} \in \mathscr{H} = 
\mathscr{V} \cup \mathscr{T} \cup \mathscr{L},~ i=1,\dots,k,
$
determines completely the expectation 
$\E \left( {\cal M}^{t}_{\bm{n}_0}|N_{t}=N \right)$.

Let us consider the case of hard-core bosons in the absence 
of magnetic fields, \textit{i.e.} the case for which $\mathscr{L}=\{1\}$.
As we will prove, after the integration of the stochastic times, 
the conditional expectation 
$\E \left( {\cal M}^{t}_{\bm{n}_0}|N_{t}=N \right)$ becomes 
\begin{eqnarray}
\label{AVERAGE0}
\E \left( \mathcal{M}^t_{\bm{n}_0} |N_{t}=N \right) = 
\media{
\mathcal{W}_N(t)
\prod_{T \in \mathscr{T}} T^{N_T}},
\end{eqnarray}
where $\mathcal{W}_N(t)$, called weight, is a function that depends only
on the multiplicities of the potential (\ref{MULTIPLICITY-V}), whereas
the other factor is a purely kinetic function that depends only on
multiplicities (\ref{MULTIPLICITY-T}).

In order to understand the behavior of Eq. (\ref{AVERAGE0}), 
and then of Eq. (\ref{TheFormulab}), let us start
to consider a non interacting system with $\hat{V}\equiv 0$.
In this case we have $\mathscr{V}=\{0\}$ 
and the weights have the simple exact expression
$\mathcal{W}_{N}^{(0)}(t)=\epsilon^N t^{N}/N!$.
Therefore for $\E \left( {\cal M}^{t}_{\bm{n}_0}|N_{t}=N \right)$ 
we are left with a residual canonical average over
the variables $N_T$ and we get  
\begin{eqnarray}
\label{E00}
\E \left( {\cal M}^{t}_{\bm{n}_0}|N_{t}=N \right)=\frac{(\epsilon t)^N}{N!}
\media{e^{\sum_{T}\log(T)N_T}} \sim \frac{(a\epsilon t)^N}{N!},
\end{eqnarray}
where $a$ is a suitable value coming from the integration over $N_T$.

The result in Eq. (\ref{E00}) can be immediately obtained by the rough estimate
$\media{ \exp [\sum_{T}\log(T)N_T]} \sim \mathrm{const}^{N}$,
based on the bounds $\exp [N\log(T_{\mathrm{min}})]
\leq\media{ \exp [\sum_{T}\log(T)N_T]}\leq \exp [N\log(T_{\mathrm{max}})]$, 
which in turn follow from the normalization $\sum_{T\in \mathscr{T}}N_T=N$.
By using a cumulant expansion theorem \cite{SHY}, however, 
it can be shown that this result becomes exact for $N \to \infty$ 
if we assume the existence of the following
rescaled cumulants of the variables $N_T$
\begin{eqnarray}
\label{CUMULANTS00}
\Sigma_{T_1,\ldots,T_k}^{(k)} =
\lim_{N \rightarrow \infty} \frac{1}{N}
\mediac{N_{T_1}\dots N_{T_k}},
\end{eqnarray}
where with $\mediac{\cdot}$ we indicate the cumulants, 
or connected correlation functions, sampled with respect to the measure 
induced by the transition matrix (\ref{MARKOV}).
The connection between these cumulants and the statistical moments introduced 
above, is well known.
For $k=1$ we have $\mediac{N_{T_1}}=\media{N_{T_1}}$,
for $k=2$ $\mediac{N_{T_1}N_{T_2}}=
\media{N_{T_1}N_{T_2}}-\media{N_{T_1}}\media{N_{T_2}}$, 
and so on.

Since the Markov chain with transition matrix (\ref{MARKOV}) 
is finite and irreducible, for $k=1$
the existence of the limit (\ref{CUMULANTS00}) is assured
by an ergodic theorem \cite{BREMAUD}. 
On the other hand, for $k>1$, the existence and the finiteness of the rescaled 
cumulants (\ref{CUMULANTS00}),
or of the more general limits (\ref{CUMULANTS0}), follows
from the exponential decay of the correlations of local functions of the
configurations along the Markov chain.
For the $T$ variables, this property amounts to say that for any $k$ 
do exist positive constants $b_k$ and $N_c(k)$ such that for any $N$
\begin{eqnarray}
\label{CUMULANTSDECAY}
\left|
\mediac{\chi_{_{T_1}}(\bm{n}_{h_1})\dots \chi_{_{T_k}}(\bm{n}_{h_k})}
\right|
\leq b_k 
\exp \left( -\frac{h_{\mathrm{max}}-h_{\mathrm{min}}} {N_c(k)} \right),
\end{eqnarray}
where $\chi_{_{T}}(\bm{n})=\delta_{T(\bm{n}),T}$ is the 
characteristic function
of the states $T$ and $h_{\mathrm{max}}$ and $h_{\mathrm{min}}$ are the maximum
and the minimum of the indices $0\leq h_l \leq N$, $l=1,\dots,k$ 
along the chain.
A numerical check of Eq.~(\ref{CUMULANTSDECAY}) for $k=2$ will be
shown in Section \ref{nresults}.

Since the rightmost expression in Eq.(\ref{E00}) has, as a function of $N$, 
an exponentially leading maximum at $N=\epsilon at$, 
we see that for $t$ large an important consequence follows. 
The full expectation
$\E \left( {\cal M}^{t}_{\bm{n}_0} \right) =
\sum_{N=0}^{\infty}\E \left( {\cal M}^{t}_{\bm{n}_0}|N_{t}=N \right)$
takes exponentially leading contributions from terms 
with $N \sim \epsilon at$.
Therefore, for $t \to \infty$ we have that 
\textit{i)} the saddle point technique used to evaluate 
the canonical averages over $N_T$ becomes exact 
and 
\textit{ii)} a further saddle point integration can be used to 
exactly resum the full expectation 
$\E \left( {\cal M}^{t}_{\bm{n}_0} \right)$.
In conclusion, we obtain an exact expression for the  
ground-state energy, namely
$\lim_{t \to \infty} -\partial_t \log 
\E \left( \mathcal{M}^t_{\bm{n}_0} \right)$.

For $\hat{V}\neq 0$, the above described scenario remains essentially 
unchanged.
In fact, even though in this case the integration of the stochastic times 
cannot be done exactly, it can be performed by a saddle point approximation, 
which becomes asymptotically exact for $N \rightarrow \infty$. 
Independently of their exact value, as we will prove later, the weights are 
bounded by 
$\mathcal{W}_{N}^{(0)}(t)e^{-V_{\mathrm{max}}t}\leq\mathcal{W}_{N}(t)
\leq\mathcal{W}_{N}^{(0)}(t) e^{-V_{\mathrm{min}}t}$ 
and behave, as a function of $N$, similarly to the
non interacting case.
The conditional expectations (\ref{AVERAGE0}), 
which now reduce to residual canonical averages over both the variables 
$N_T$ and $N_V$, can be evaluated analogously to Eq. (\ref{E00}). 
The result again implies that the saddle point integrations 
we perform to evaluate 
\textit{i)} the weights, 
\textit{ii)} the residual canonical averages,
and \textit{iii)} the sum of the series
$\sum_{N=0}^{\infty}\E \left( {\cal M}^{t}_{\bm{n}_0}|N_{t}=N \right)$
all become exact in the limit $t \rightarrow \infty$.
The conclusion is that
for hard-core boson systems in the absence of magnetic fields,
\textit{i.e.} $\mathscr{L}=\{1\}$, the ground-state energy $E_{0B}$ 
is determined by the following scalar equation involving all the cumulants
of the Markov dynamics
\begin{eqnarray}
\label{E0BEQ0}
\sum_{k=1}^{\infty}
\frac{1}{k!}
\sum_{\alpha_1\in\mathscr{H}}
\ldots
\sum_{\alpha_k\in\mathscr{H}}
\Sigma_{\alpha_1,\dots,\alpha_k}^{(k)}
u_{\alpha_1}(E_{0B}) \dots u_{\alpha_k}(E_{0B}) =0 ,
\end{eqnarray}
where $\mathscr{H} = \mathscr{V}\cup\mathscr{T}$, 
and the vector $\bm{u}^\mathrm{T} = (\dots u_V \dots ;\dots u_T \dots)$
with $V \in \mathscr{V}$ and $T \in \mathscr{T}$ is defined as 
\begin{eqnarray}
\bm{u}^\mathrm{T} (E_{0B})
= (\ldots -\log[(-E_{0B}+V)/\epsilon] \ldots;
\ldots \log T \ldots).
\end{eqnarray}
The asymptotic rescaled cumulants $\Sigma_{\alpha_1,\ldots,\alpha_k}^{(k)}$
are defined as
\begin{eqnarray}
\label{CUMULANTS0}
\Sigma_{\alpha_1,\ldots,\alpha_k}^{(k)} =
\lim_{N \rightarrow \infty} \frac{\mediac{N_{\alpha_1}\dots N_{\alpha_k}}}{N},
\end{eqnarray}
where with $\mediac{\cdot}$ we mean the cumulants,
or connected correlation functions, of the multiplicities $N_\alpha$,
$\alpha \in \mathscr{H}$, 
sampled with respect to the measure induced at $N$ jumps by the Markov chain 
with transition matrix (\ref{MARKOV}).

For $\hat{V}\equiv 0$, Eq. (\ref{E0BEQ0}) can be solved explicitly
and we get the following exact expression for the ground-state energy 
$E_{0B}^{(0)}$ of a non interacting hard-core boson system
\begin{eqnarray}
\label{E000}
E_{0B}^{(0)} = -\epsilon \exp\left[\sum_{k=1}^{\infty}\frac{1}{k!}
\sum_{T_1 \in\mathscr{T}}
\ldots 
\sum_{T_k \in\mathscr{T}}
\Sigma^{(k)}_{T_1,\ldots,T_k}
\log T_1 \dots \log T_k
\right].
\end{eqnarray}
For a generic $\hat{V}$, 
Eq.~(\ref{E0BEQ0}), with the series truncated at an
arbitrary finite order $k_\mathrm{max}$, 
can be solved numerically in a straightforward way by 
using the bounds  $E_{0B}^{(0)} \leq  E_{0B} \leq 0$.

Equation (\ref{E0BEQ0}) allows,
via the Hellman-Feynman theorem (\ref{HF}), 
also to evaluate the expectation of any other operator 
in the ground state of the chosen Hamiltonian
(see Ref.~\cite{OP1} for more details).
Moreover, it is clear that the cumulants $\bm{\Sigma}^{(k)}$ 
depend only on the structure of the Hamiltonian operator, 
not on the values of the Hamiltonian parameters.
Therefore, once the $\bm{\Sigma}^{(k)}$ are known,  
all the evaluated ground-state expectations are analytical 
functions of the Hamiltonian parameters.

Our analytical formula for the ground-state energy (\ref{E0BEQ0})
rests on the knowledge of the asymptotic rescaled cumulants.
In some special cases, it could be possible to evaluate, 
at least approximatively, also these cumulants analytically.
We refer to Section \ref{nresults} for a detailed discussion
on how determining the cumulants in a numerical way.
In that Section, some example cases are worked out explicitly.
Finally, we refer to Section \ref{conclusions} for a scenario of possible 
applications that exploit the analytic character of our approach.

\section{Probabilistic expectation in the long-time limit: 
proof of Eqs. (\ref{E0BEQ0}-\ref{E000})} 
\label{pelt}
 
\subsection{Canonical decomposition of the expectation}
\label{decomposition}
To evaluate the expectation  
$\E \left( {\cal M}^{t}_{\bm{n}_0} \right)$, 
we decompose it as a series of conditional expectations 
with a fixed number of jumps (canonical averages)
\begin{eqnarray}
\label{EXPANSION}
\E \left( {\cal M}^{t}_{\bm{n}_0} \right) &=&
\sum_{N=0}^{\infty}\E \left( {\cal M}^{t}_{\bm{n}_0}|N_{t}=N \right). 
\end{eqnarray}
In the canonical averages the stochastic jump times are easily 
integrated out.
In fact, multiplying the stochastic functional (\ref{FORMULA D})
with the condition $N_t=N$ by the infinitesimal probability, 
\begin{eqnarray*}\fl
dP_N =
e^{-|\Gamma| s_1 }\rho ds_{1} ~
e^{-|\Gamma| \left( s_2-s_1 \right) }\rho ds_{2} ~
\ldots 
e^{-|\Gamma| \left( s_N-s_{N-1} \right) } \rho ds_{N}~
e^{-|\Gamma| \left( t-s_N \right) },
\end{eqnarray*}
to have the jumps $1,2,\ldots,N$ of the independent Poisson processes 
in the intervals 
$[s_1,s_1+ds_1), [s_2,s_2+ds_2), \ldots, [s_N,s_N+ds_N)$, respectively, 
and integrating over the times $s_1,s_2, \ldots, s_N$, we get
\begin{eqnarray}
\label{canave}
\E \left( {\cal M}^{t}_{\bm{n}_0}|N_{t}=N \right) &=&
\sum_{r \in \Omega_N}
\mathcal{S}_N^{(r)} \mathcal{K}_N^{(r)} \mathcal{W}_N^{(r)}(t),
\end{eqnarray}
where $\Omega_N=\Omega_N(\bm{n}_{0})$ is the set of 
all possible trajectories with $N$ jumps branching from the initial 
configuration $\bm{n}_{0}$ and 
\begin{eqnarray}
\label{signs}
\mathcal{S}_N^{(r)} &=&
\lambda_1^{(r)} \lambda_2^{(r)} \ldots \lambda_N^{(r)} ,
\end{eqnarray} 
\begin{eqnarray}
\label{kappas}
\mathcal{K}_N^{(r)} &=& \epsilon^{-N}
\eta_1^{(r)} \eta_2^{(r)} \ldots \eta_N^{(r)} ,
\end{eqnarray} 
\begin{eqnarray}\fl
\label{weights}
\mathcal{W}^{(r)}_N(t) =
\epsilon^{N} \int_{0}^{t} ds_{1} 
\int_{s_{1}}^{t} ds_{2}
\dots 
\int_{s_{N-1}}^{t} ds_{N}
e^{-V_{0}s_{1}
-V_1^{(r)} \left( s_{2}-s_{1} \right)
\dots 
-V_N^{(r)} \left( t-s_{N} \right) }.
\end{eqnarray} 
Equations (\ref{signs}), (\ref{kappas}) and (\ref{weights}) define
three dimensionless quantities related, respectively, 
to the sequences of the signs (more generally, of the phases), 
of the hopping parameters and of the potential values 
associated to the $r$th trajectory with $N>0$ jumps.
We have $\mathcal{S}_0^{(r)}=\mathcal{K}_0^{(r)}=1$ 
and $\mathcal{W}^{(r)}_0(t) = e^{-V_0t}$ in the case $N=0$.
The quantities $\mathcal{W}^{(r)}_N(t)$ are positive definite 
and will be called weights in the following.

From Eq.~(\ref{signs}), it is clear that only the trajectories
formed by a sequence of active links contribute to the sum 
in Eq.~(\ref{canave}). 
Hereafter, therefore, we restrict $\Omega_N$ 
to be the set of these effective trajectories with $N$ jumps
branching from $\bm{n}_0$ and we exclude the value $\lambda=0$ 
from the set $\mathscr{L}$.  

The sum over the set $\Omega_N$ can be expressed as an average, 
$\media{\cdot}$, over the trajectories with $N$ jumps generated
by extracting with uniform probability one of the active links 
available at the configurations 
$\bm{n}_0,\bm{n}_1,\ldots,\bm{n}_{N-1}$.
The probability associated to the $r$th trajectory generated in this way
is $p_N^{(r)} = \prod_{k=0}^{N-1} 1/{A_k^{(r)}}$ and we have
\begin{eqnarray}
\label{SUM-AVERAGE}
\sum_{r \in \Omega_N} \mathcal{S}_N^{(r)} \mathcal{K}_N^{(r)}
\mathcal{W}_N^{(r)}(t) &=& 
\sum_{r \in \Omega_N} 
p_N^{(r)} \mathcal{S}_N^{(r)} \mathcal{K}_N^{(r)} \mathcal{W}_N^{(r)}(t)
\prod_{k=0}^{N-1} A_k^{(r)}
\nonumber\\
&=& \media{
\mathcal{S}_N
\mathcal{W}_N(t)
\prod_{k=0}^{N-1} T_k},
\end{eqnarray}
where 
$\media{\cdot} = \sum_{r \in \Omega_N} \cdot~ p_N^{(r)}$.
Note that 
$\sum_{r \in \Omega_N} p_N^{(r)} = 1$.
By using the definitions (\ref{MULTIPLICITY-L}) and (\ref{MULTIPLICITY-T}),
we rewrite the canonical averages as
\begin{eqnarray}
\label{AVERAGES}
\E \left( \mathcal{M}^t_{\bm{n}_0} |N_{t}=N \right) = 
\media{
\mathcal{W}_N(t)
\prod_{\lambda \in \mathscr{L}} \lambda^{N_\lambda}
\prod_{T \in \mathscr{T}} T^{N_T}}.
\end{eqnarray}
In the following, we will consider only hard-core bosons in the absence 
of magnetic fields. 
In this case, $\mathscr{L}=\{ 1\}$ and we are left with averages 

of positive definite quantities
\begin{eqnarray}
\label{AVERAGE}
\E \left( \mathcal{M}^t_{\bm{n}_0} |N_{t}=N \right) = 
\media{
\mathcal{W}_N(t)
\prod_{T \in \mathscr{T}} T^{N_T}}.
\end{eqnarray}

\subsection{Evaluation of the weights}
\label{evaluateweights}
In this Section, first we find a recursive differential 
equation for the weights $\mathcal{W}_{N}(t)$. 
Then, taking the Laplace transform of this equation,
we realize that the weights depend only on the multiplicities $N_{V}$ 
of the potential, not on the detailed sequence $V_0,V_1, \ldots, V_N$.
Finally, we use a saddle-point technique on the complex plane
to obtain an explicit expression for the weights,
which becomes exact in the limit $N \to \infty$.
 
Equation (\ref{weights}), 
which defines the weights $\mathcal{W}_{N}(t)$ for $N \geq 1$,
can be rewritten as (for simplicity, we omit the trajectory index $(r)$ 
but we stress the dependence on the sequence of the potential values)
\begin{eqnarray*}
\mathcal{W}_{N}(t;V_0,V_1,\dots,V_N) =
G_{N}(t;V_0,V_1,\dots,V_N) e^{-V_{N}t},
\end{eqnarray*}
where 
\begin{eqnarray*}\fl
\label{G}
G_{N}(t;V_0,V_1,\dots,V_N) &=& \epsilon^N
\int_{0}^{t} ds_1 
\int_{0}^{t} ds_2 \ldots 
\int_{0}^{t} ds_N 
\theta(s_{2}-s_{1}) \theta(s_{3}-s_{2}) \dots 
\\ &&\times 
\theta(s_{N}-s_{N-1}) 
\exp\left(\Delta_{1}s_{1}+\Delta_{2}s_{2}+ \dots +\Delta_{N}s_{N}\right),
\end{eqnarray*}
with $\Delta_{k}=V_{k}-V_{k-1}$, $ k=1,2, \dots ,N$.
By evaluating the derivative of $G_{N}(t)$, for $N >0$, 
with respect to $t$,
\begin{eqnarray*}
\partial_{t} G_{N}(t;V_0,V_1,\dots,V_N) 
= \epsilon  
G_{N-1}(t;V_0,V_1,\dots,V_{N-1}) 
\exp \left(\Delta_{N}t\right),
\end{eqnarray*} 
where $G_0(t;V_0)=1$,
we find the following recursive ordinary differential equation
for $\mathcal{W}_{N}(t)$
\begin{eqnarray}
\label{WDOT}
\partial_{t} \mathcal{W}_{N}(t;V_0,V_1,\dots,V_N) &=&
\epsilon \mathcal{W}_{N-1}(t;V_0,V_{1},\dots,V_{N-1}) 
\nonumber \\ &&
-V_{N}\mathcal{W}_{N}(t;V_0,V_1,\dots,V_N). 
\end{eqnarray}

Since Eq.~(\ref{WDOT}) is linear in $\mathcal{W}_{N}(t)$ and 
$\mathcal{W}_{N-1}(t)$, it is convenient to introduce the Laplace transform 
\begin{eqnarray}
\widetilde{\mathcal{W}}_{N}(z) = 
\int_{0}^{\infty}dt e^{-zt}{\mathcal{W}}_{N}(t), 
\quad z \in \mathbb{C}
\nonumber.
\end{eqnarray}
By observing that $\mathcal{W}_N(0)=0$ for $N > 0$, 
Eq.~(\ref{WDOT}) reduces to the following recursive algebraic equation  
for $\widetilde{\mathcal{W}}_{N}(z)$
\begin{eqnarray}
z\widetilde{\mathcal{W}}_{N}(z)=\epsilon 
\widetilde{\mathcal{W}}_{N-1}(z)-V_{N}\widetilde{\mathcal{W}}_{N}(z),
\end{eqnarray}
from which we get
\begin{eqnarray}
\label{ltiter}
\widetilde{\mathcal{W}}_{N}(z) = 
\epsilon (z+V_{N})^{-1} \widetilde{\mathcal{W}}_{N-1}(z) .
\end{eqnarray}
We recall that for $N=0$ we have
$\mathcal{W}_0(t;V_0) = e^{-V_0t}$ and, therefore,
$\widetilde{\mathcal{W}}_{0}(z)= (z+V_{0})^{-1}$.
By iterating Eq.~(\ref{ltiter}), we arrive at the solution 
\begin{eqnarray*}
\label{WTILDE0}
\widetilde{\mathcal{W}}_{N}(z)=\epsilon^N\prod_{k=0}^{N}\frac{1}{z+V_{k}}, 
\end{eqnarray*}
which, in terms of the multiplicities (\ref{MULTIPLICITY-V}),
takes the form
\begin{eqnarray}
\label{WTILDE}
\widetilde{\mathcal{W}}_{N}(z) = \epsilon^{N} \prod_{V\in\mathscr{V}}
(z+V)^{-N_V}.
\end{eqnarray}
This expression shows that, for any value of $N$, 
the weights depend only on the multiplicities $N_V$, 
\textit{i.e.} $\mathcal{W}_{N}(t;V_0,V_1,\dots,V_N)
=\mathcal{W}_{N}(t;\{N_{V}\})$.

The explicit inversion of the Laplace transform (\ref{WTILDE}) 
can be done analytically only for $N$ large.
However, as we will immediately prove, this is the limit we are interested in.
In fact, in the case $\hat{V}\equiv 0$, we have
$\mathcal{W}_{N}(t)=\mathcal{W}_{N}^{(0)}(t)=\epsilon^N t^{N}/N!$, 
which is easily obtained by direct integration of Eq.~(\ref{weights}).
For $t$ large 
$\mathcal{W}_{N}^{(0)}(t)$ has, as a function of $N$, 
a maximum at $N(t)=\epsilon t$ and, around this maximum, is well
approximated by a Gaussian of width $\sqrt{N(t)}$. 
For $\hat{V}$ generic, let us indicate with $V_{\mathrm{min}}$ and 
$V_{\mathrm{max}}$ the minimum and maximum elements of $\mathscr{V}$ 
such that $N_V\neq 0$ for the trajectory we are considering.
In Eq. (\ref{weights}) we thus have 
$V_{\mathrm{min}} \leq V_k \leq V_{\mathrm{max}}$, $k=0,1,\dots,N$ and 
\begin{eqnarray}
e^{-V_{\mathrm{max}} t}
\leq
e^{-V_{0}s_{1}-V_1 \left( s_{2}-s_{1} \right)
\dots -V_N \left( t-s_{N} \right) }
\leq
e^{-V_{\mathrm{min}} t}.
\end{eqnarray} 
We conclude that the weights are bounded by 
\begin{eqnarray}
\mathcal{W}_{N}^{(0)}(t)e^{-V_{\mathrm{max}}t}\leq\mathcal{W}_{N}(t)\leq
\mathcal{W}_{N}^{(0)}(t) e^{-V_{\mathrm{min}}t} ,
\end{eqnarray}
so that, for $t \to \infty$, exponentially leading contributions
are obtained from values of $N$ in the range 
$[\epsilon t -\sqrt{\epsilon t}, \epsilon t+\sqrt{\epsilon t}]$. 

According to Eq.~(\ref{WTILDE}), $\widetilde{\mathcal{W}}_{N}(z)$ 
has $m_N \leq m_{\mathscr{V}}$ poles at the points $z^{}_{V}=-V$ such 
that $V\in\mathscr{V}$ and $N_V \neq 0$.
For $N$ sufficiently large, the number $m_N$ of these poles 
approaches $m_{\mathscr{V}}$. 
For any finite value of $N$,
the poles of $\widetilde{\mathcal{W}}_{N}(z)$ are confined in the real 
segment $[-V_{\mathrm{max}},-V_{\mathrm{min}}]$.
Recalling the rule for the Laplace inverse transformation, we have 
\begin{eqnarray}
\mathcal{W}_{N}(t)=\frac{1}{2\pi i}\int_{\Gamma} dz
e^{zt} \widetilde{\mathcal{W}}_{N}(z) \nonumber,
\end{eqnarray}
where the integration contour $\Gamma$ can be any line parallel to the 
imaginary axis and contained in the analyticity domain of 
the Laplace transform. 
In our case, $\Gamma$ must be in the domain $\Re z > -V_{\mathrm{min}}$.
By Jordan's lemma, the contour can be closed to infinity   
in the left half-plane $\Re z <-V_{\mathrm{min}}$ without changing
the integration result.
Finally, by Cauchy's theorem, $\Gamma$ can be deformed into any other 
anti-clockwise closed contour $\Gamma'$ still containing all the poles 
$z_{V}$.

By using the expression (\ref{WTILDE}) for $\widetilde{\mathcal{W}}_{N}(z)$, 
we write its antitransform as
\begin{eqnarray}
\label{LINV}
\mathcal{W}_{N}(t)=\frac{1}{2\pi i \epsilon}
\int_{\Gamma} dz \exp\left[N \varphi(z) \right],
\end{eqnarray}
where
\begin{eqnarray}
\label{PHI}
\varphi(z) =
\frac{zt}{N} - 
\sum_{V \in \mathscr{V}}\frac{N_{V}}{N} \log\frac{z+V}{\epsilon}.
\end{eqnarray}
For $t$ large and $N \sim t$, since also $N_{V} \sim N$ due to the 
ergodicity of the trajectories,
we can evaluate the complex integral (\ref{LINV}) 
by a saddle-point technique. 
Let us call $z_0$ the point of the complex plane where $\varphi(z)$ 
is stationary, \textit{i.e.} $\partial_z \varphi(z_0)=0$.
We deform the contour $\Gamma$ into a new one $\Gamma'$ (steepest descent)
such that, for $z \in \Gamma'$, $\Re \varphi(z)$ has a maximum at $z_{0}$, 
whereas $\Im \varphi(z)$ is constant, at least at the first order 
in its Taylor expansion around $z_{0}$.
Provided that the contour $\Gamma'$ remains in the analyticity
domain of $\varphi(z)$, the following standard result applies~\cite{BRUIJN} 
\begin{eqnarray}
\label{SADDLEPOINT}
\mathcal{W}_{N}(t) = \frac{1}{2\pi i \epsilon}
\sqrt{
\frac{2\pi}{N \left\vert \partial_z^2 \varphi(z_{0}) \right\vert} }
\exp\left[ i\alpha+N\varphi(z_{0})  \right], 
\end{eqnarray}
where $\alpha$ is defined as 
\begin{eqnarray}
\alpha=\frac{\pi}{2}-\frac{1}{2}
\Arg \partial_z^2 \varphi(z_{0})
\end{eqnarray}
and represents the angle formed, at the saddle point $z_0$, 
by the new contour $\Gamma'$ with respect to the original one $\Gamma$.

Now we calculate the first and second derivatives of $\varphi(z)$ 
and the saddle point $z_{0}$.
From Eq.~(\ref{PHI}) we have
\begin{eqnarray}
\label{DPHI}
\partial_z \varphi(z) &=&
\frac{t}{N}-\sum_{V \in \mathscr{V}}\frac{N_{V}}{N}
\frac{1}{z+V},
\\
\label{DDPHI}
\partial_z^2 \varphi(z) &=&
\sum_{V \in \mathscr{V}}\frac{N_{V}}{N}
\frac{1}{(z+V)^{2}}.
\end{eqnarray}
In terms of $x_{0}=\Re z_0$ and $y_{0}=\Im z_0 $, 
the saddle point equation, $\partial_z \varphi(z_{0})=0$, reads
\numparts
\begin{eqnarray}
\frac{t}{N}-
\sum_{V \in \mathscr{V}}\frac{N_{V}}{N}
\frac{x_{0}}{(x_{0}+V)^{2}+y_{0}^{2}} = 0 , 
\label{ReSYSTEM}
\\
\sum_{V \in \mathscr{V}}\frac{N_{V}}{N}
\frac{y_{0}}{(x_{0}+V)^{2}+y_{0}^{2}} = 0.
\label{ImSYSTEM}
\end{eqnarray}
\endnumparts
As $N_{V} \geq 0$, Eq.~(\ref{ImSYSTEM}) is satisfied only by $y_{0}=0$. 
Hence, we are left with the following equation for $x_{0}$
\begin{eqnarray}
\label{X}
\sum_{V \in \mathscr{V}}\frac{N_{V}}{x_{0}+V}=t .
\end{eqnarray}
For any $t>0$, Eq.~(\ref{X}) has $m_N$ solutions.
The first $m_N-1$ solutions, 
ordered according to their increasing value, 
are in the range $-V_{\mathrm{max}} < x_0 <-V_{\mathrm{min}}$, 
whereas the last one is such that $x_0 > -V_{\mathrm{min}}$.
This is the only solution compatible with the condition that the
integration contour $\Gamma$, passing through $z_0=x_0+iy_0$, 
is contained in analyticity region of $\varphi(z)$.
Therefore, for any $t>0$, we have one and only one saddle point determined
by Eq.~(\ref{X}) with the constraint $x_0 > -V_{\mathrm{min}}$. 

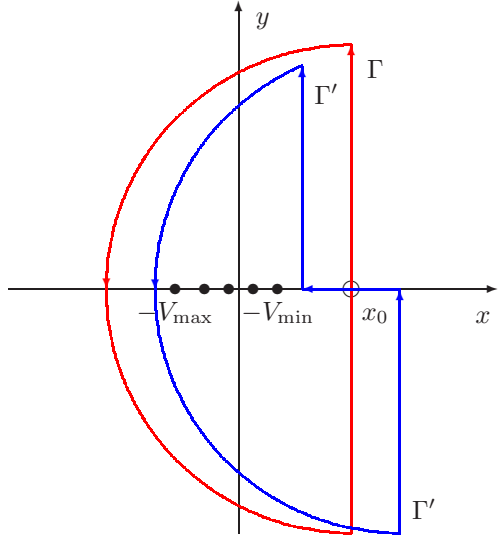
\begin{figure}[t]
\centering
\unitlength 0.65mm
\begin{picture}(100.00,110.00)(0,0)
\linethickness{0.15mm}
\put(0.00,50.00){\line(1,0){98.00}}
\put(100.00,50.00){\vector(1,0){0.}}
\linethickness{0.15mm}
\put(47.00,0.00){\line(0,1){107.00}}
\put(47.00,109.00){\vector(0,1){0.}}
\linethickness{0.25mm}
\put(97.00,43.00){\makebox(0.00,0.00)[cb]{$x$}}
\put(52.00,105){\makebox(0.00,0.00)[cc]{$y$}}
\put(75.00,43.00){\makebox(0.00,0.00)[cb]{$x_0$}}
\put(70.00,50.00){\circle{3.00}}
\put(75.00,95){\makebox(0.00,0.00)[cc]{$\Gamma$}}
\put(65.00,90){\makebox(0.00,0.00)[cc]{$\Gamma'$}}
\put(85.00,5){\makebox(0.00,0.00)[cc]{$\Gamma'$}}
\put(55.00,50.00){\circle*{2.00}}
\put(55.00,43.00){\makebox(0.00,0.00)[cb]{$-V_{\mathrm{min}}$}}
\put(50.00,50.00){\circle*{2.00}}
\put(45.00,50.00){\circle*{2.00}}
\put(40.00,50.00){\circle*{2.00}}
\put(34.00,50.00){\circle*{2.00}}
\put(34.00,43.00){\makebox(0.00,0.00)[cb]{$-V_{\mathrm{max}}$}}
\linethickness{0.25mm}
\put(70.00,0.00){\textcolor{red}{\line(0,1){100.00}}}
\put(70.00,100.00){\textcolor{red}{\vector(0,1){0.}}}
\multiput(69.00,100.00)(0.99,-0.02){1}{\textcolor{red}{\line(1,0){0.99}}}
\multiput(68.51,99.98)(0.99,0.02){1}{\textcolor{red}{\line(1,0){0.99}}}
\multiput(67.52,99.94)(0.99,0.04){1}{\textcolor{red}{\line(1,0){0.99}}}
\multiput(66.52,99.88)(0.99,0.06){1}{\textcolor{red}{\line(1,0){0.99}}}
\multiput(65.53,99.80)(0.99,0.08){1}{\textcolor{red}{\line(1,0){0.99}}}
\multiput(64.54,99.70)(0.99,0.10){1}{\textcolor{red}{\line(1,0){0.99}}}
\multiput(63.56,99.58)(0.99,0.12){1}{\textcolor{red}{\line(1,0){0.99}}}
\multiput(62.57,99.45)(0.98,0.14){1}{\textcolor{red}{\line(1,0){0.98}}}
\multiput(61.59,99.29)(0.98,0.16){1}{\textcolor{red}{\line(1,0){0.98}}}
\multiput(60.61,99.11)(0.98,0.18){1}{\textcolor{red}{\line(1,0){0.98}}}
\multiput(59.64,98.91)(0.49,0.10){2}{\textcolor{red}{\line(1,0){0.49}}}
\multiput(58.67,98.70)(0.49,0.11){2}{\textcolor{red}{\line(1,0){0.49}}}
\multiput(57.70,98.46)(0.48,0.12){2}{\textcolor{red}{\line(1,0){0.48}}}
\multiput(56.74,98.21)(0.48,0.13){2}{\textcolor{red}{\line(1,0){0.48}}}
\multiput(55.78,97.94)(0.48,0.14){2}{\textcolor{red}{\line(1,0){0.48}}}
\multiput(54.83,97.64)(0.48,0.15){2}{\textcolor{red}{\line(1,0){0.48}}}
\multiput(53.89,97.33)(0.31,0.10){3}{\textcolor{red}{\line(1,0){0.31}}}
\multiput(52.95,97.00)(0.31,0.11){3}{\textcolor{red}{\line(1,0){0.31}}}
\multiput(52.02,96.66)(0.31,0.12){3}{\textcolor{red}{\line(1,0){0.31}}}
\multiput(51.10,96.29)(0.31,0.12){3}{\textcolor{red}{\line(1,0){0.31}}}
\multiput(50.18,95.90)(0.31,0.13){3}{\textcolor{red}{\line(1,0){0.31}}}
\multiput(49.27,95.50)(0.30,0.13){3}{\textcolor{red}{\line(1,0){0.30}}}
\multiput(48.37,95.08)(0.23,0.11){4}{\textcolor{red}{\line(1,0){0.23}}}
\multiput(47.48,94.64)(0.22,0.11){4}{\textcolor{red}{\line(1,0){0.22}}}
\multiput(46.59,94.18)(0.22,0.11){4}{\textcolor{red}{\line(1,0){0.22}}}
\multiput(45.72,93.71)(0.22,0.12){4}{\textcolor{red}{\line(1,0){0.22}}}
\multiput(44.86,93.22)(0.22,0.12){4}{\textcolor{red}{\line(1,0){0.22}}}
\multiput(44.00,92.71)(0.21,0.13){4}{\textcolor{red}{\line(1,0){0.21}}}
\multiput(43.16,92.18)(0.21,0.13){4}{\textcolor{red}{\line(1,0){0.21}}}
\multiput(42.32,91.64)(0.17,0.11){5}{\textcolor{red}{\line(1,0){0.17}}}
\multiput(41.50,91.08)(0.16,0.11){5}{\textcolor{red}{\line(1,0){0.16}}}
\multiput(40.69,90.51)(0.16,0.11){5}{\textcolor{red}{\line(1,0){0.16}}}
\multiput(39.89,89.92)(0.16,0.12){5}{\textcolor{red}{\line(1,0){0.16}}}
\multiput(39.10,89.31)(0.16,0.12){5}{\textcolor{red}{\line(1,0){0.16}}}
\multiput(38.33,88.69)(0.16,0.12){5}{\textcolor{red}{\line(1,0){0.16}}}
\multiput(37.57,88.05)(0.15,0.13){5}{\textcolor{red}{\line(1,0){0.15}}}
\multiput(36.82,87.40)(0.15,0.13){5}{\textcolor{red}{\line(1,0){0.15}}}
\multiput(36.08,86.73)(0.12,0.11){6}{\textcolor{red}{\line(1,0){0.12}}}
\multiput(35.35,86.05)(0.12,0.11){6}{\textcolor{red}{\line(1,0){0.12}}}
\multiput(34.64,85.36)(0.12,0.12){6}{\textcolor{red}{\line(1,0){0.12}}}
\multiput(33.95,84.65)(0.12,0.12){6}{\textcolor{red}{\line(0,1){0.12}}}
\multiput(33.27,83.92)(0.11,0.12){6}{\textcolor{red}{\line(0,1){0.12}}}
\multiput(32.60,83.18)(0.11,0.12){6}{\textcolor{red}{\line(0,1){0.12}}}
\multiput(31.95,82.43)(0.13,0.15){5}{\textcolor{red}{\line(0,1){0.15}}}
\multiput(31.31,81.67)(0.13,0.15){5}{\textcolor{red}{\line(0,1){0.15}}}
\multiput(30.69,80.90)(0.12,0.16){5}{\textcolor{red}{\line(0,1){0.16}}}
\multiput(30.08,80.11)(0.12,0.16){5}{\textcolor{red}{\line(0,1){0.16}}}
\multiput(29.49,79.31)(0.12,0.16){5}{\textcolor{red}{\line(0,1){0.16}}}
\multiput(28.92,78.50)(0.11,0.16){5}{\textcolor{red}{\line(0,1){0.16}}}
\multiput(28.36,77.68)(0.11,0.16){5}{\textcolor{red}{\line(0,1){0.16}}}
\multiput(27.82,76.84)(0.11,0.17){5}{\textcolor{red}{\line(0,1){0.17}}}
\multiput(27.29,76.00)(0.13,0.21){4}{\textcolor{red}{\line(0,1){0.21}}}
\multiput(26.78,75.14)(0.13,0.21){4}{\textcolor{red}{\line(0,1){0.21}}}
\multiput(26.29,74.28)(0.12,0.22){4}{\textcolor{red}{\line(0,1){0.22}}}
\multiput(25.82,73.41)(0.12,0.22){4}{\textcolor{red}{\line(0,1){0.22}}}
\multiput(25.36,72.52)(0.11,0.22){4}{\textcolor{red}{\line(0,1){0.22}}}
\multiput(24.92,71.63)(0.11,0.22){4}{\textcolor{red}{\line(0,1){0.22}}}
\multiput(24.50,70.73)(0.11,0.23){4}{\textcolor{red}{\line(0,1){0.23}}}
\multiput(24.10,69.82)(0.13,0.30){3}{\textcolor{red}{\line(0,1){0.30}}}
\multiput(23.71,68.90)(0.13,0.31){3}{\textcolor{red}{\line(0,1){0.31}}}
\multiput(23.34,67.98)(0.12,0.31){3}{\textcolor{red}{\line(0,1){0.31}}}
\multiput(23.00,67.05)(0.12,0.31){3}{\textcolor{red}{\line(0,1){0.31}}}
\multiput(22.67,66.11)(0.11,0.31){3}{\textcolor{red}{\line(0,1){0.31}}}
\multiput(22.36,65.17)(0.10,0.31){3}{\textcolor{red}{\line(0,1){0.31}}}
\multiput(22.06,64.22)(0.15,0.48){2}{\textcolor{red}{\line(0,1){0.48}}}
\multiput(21.79,63.26)(0.14,0.48){2}{\textcolor{red}{\line(0,1){0.48}}}
\multiput(21.54,62.30)(0.13,0.48){2}{\textcolor{red}{\line(0,1){0.48}}}
\multiput(21.30,61.33)(0.12,0.48){2}{\textcolor{red}{\line(0,1){0.48}}}
\multiput(21.09,60.36)(0.11,0.49){2}{\textcolor{red}{\line(0,1){0.49}}}
\multiput(20.89,59.39)(0.10,0.49){2}{\textcolor{red}{\line(0,1){0.49}}}
\multiput(20.71,58.41)(0.18,0.98){1}{\textcolor{red}{\line(0,1){0.98}}}
\multiput(20.55,57.43)(0.16,0.98){1}{\textcolor{red}{\line(0,1){0.98}}}
\multiput(20.42,56.44)(0.14,0.98){1}{\textcolor{red}{\line(0,1){0.98}}}
\multiput(20.30,55.46)(0.12,0.99){1}{\textcolor{red}{\line(0,1){0.99}}}
\multiput(20.20,54.47)(0.10,0.99){1}{\textcolor{red}{\line(0,1){0.99}}}
\multiput(20.12,53.48)(0.08,0.99){1}{\textcolor{red}{\line(0,1){0.99}}}
\multiput(20.06,52.48)(0.06,0.99){1}{\textcolor{red}{\line(0,1){0.99}}}
\multiput(20.02,51.49)(0.04,0.99){1}{\textcolor{red}{\line(0,1){0.99}}}
\multiput(20.00,50.50)(0.02,0.99){1}{\textcolor{red}{\line(0,1){0.99}}}
\put(20.00,50.00){\textcolor{red}{\vector(0,-1){0.}}}
\put(20.00,49.50){\textcolor{red}{\line(0,1){0.99}}}
\multiput(20.00,49.50)(0.02,-0.99){1}{\textcolor{red}{\line(0,-1){0.99}}}
\multiput(20.02,48.51)(0.04,-0.99){1}{\textcolor{red}{\line(0,-1){0.99}}}
\multiput(20.06,47.52)(0.06,-0.99){1}{\textcolor{red}{\line(0,-1){0.99}}}
\multiput(20.12,46.52)(0.08,-0.99){1}{\textcolor{red}{\line(0,-1){0.99}}}
\multiput(20.20,45.53)(0.10,-0.99){1}{\textcolor{red}{\line(0,-1){0.99}}}
\multiput(20.30,44.54)(0.12,-0.99){1}{\textcolor{red}{\line(0,-1){0.99}}}
\multiput(20.42,43.56)(0.14,-0.98){1}{\textcolor{red}{\line(0,-1){0.98}}}
\multiput(20.55,42.57)(0.16,-0.98){1}{\textcolor{red}{\line(0,-1){0.98}}}
\multiput(20.71,41.59)(0.18,-0.98){1}{\textcolor{red}{\line(0,-1){0.98}}}
\multiput(20.89,40.61)(0.10,-0.49){2}{\textcolor{red}{\line(0,-1){0.49}}}
\multiput(21.09,39.64)(0.11,-0.49){2}{\textcolor{red}{\line(0,-1){0.49}}}
\multiput(21.30,38.67)(0.12,-0.48){2}{\textcolor{red}{\line(0,-1){0.48}}}
\multiput(21.54,37.70)(0.13,-0.48){2}{\textcolor{red}{\line(0,-1){0.48}}}
\multiput(21.79,36.74)(0.14,-0.48){2}{\textcolor{red}{\line(0,-1){0.48}}}
\multiput(22.06,35.78)(0.15,-0.48){2}{\textcolor{red}{\line(0,-1){0.48}}}
\multiput(22.36,34.83)(0.10,-0.31){3}{\textcolor{red}{\line(0,-1){0.31}}}
\multiput(22.67,33.89)(0.11,-0.31){3}{\textcolor{red}{\line(0,-1){0.31}}}
\multiput(23.00,32.95)(0.12,-0.31){3}{\textcolor{red}{\line(0,-1){0.31}}}
\multiput(23.34,32.02)(0.12,-0.31){3}{\textcolor{red}{\line(0,-1){0.31}}}
\multiput(23.71,31.10)(0.13,-0.31){3}{\textcolor{red}{\line(0,-1){0.31}}}
\multiput(24.10,30.18)(0.13,-0.30){3}{\textcolor{red}{\line(0,-1){0.30}}}
\multiput(24.50,29.27)(0.11,-0.23){4}{\textcolor{red}{\line(0,-1){0.23}}}
\multiput(24.92,28.37)(0.11,-0.22){4}{\textcolor{red}{\line(0,-1){0.22}}}
\multiput(25.36,27.48)(0.11,-0.22){4}{\textcolor{red}{\line(0,-1){0.22}}}
\multiput(25.82,26.59)(0.12,-0.22){4}{\textcolor{red}{\line(0,-1){0.22}}}
\multiput(26.29,25.72)(0.12,-0.22){4}{\textcolor{red}{\line(0,-1){0.22}}}
\multiput(26.78,24.86)(0.13,-0.21){4}{\textcolor{red}{\line(0,-1){0.21}}}
\multiput(27.29,24.00)(0.13,-0.21){4}{\textcolor{red}{\line(0,-1){0.21}}}
\multiput(27.82,23.16)(0.11,-0.17){5}{\textcolor{red}{\line(0,-1){0.17}}}
\multiput(28.36,22.32)(0.11,-0.16){5}{\textcolor{red}{\line(0,-1){0.16}}}
\multiput(28.92,21.50)(0.11,-0.16){5}{\textcolor{red}{\line(0,-1){0.16}}}
\multiput(29.49,20.69)(0.12,-0.16){5}{\textcolor{red}{\line(0,-1){0.16}}}
\multiput(30.08,19.89)(0.12,-0.16){5}{\textcolor{red}{\line(0,-1){0.16}}}
\multiput(30.69,19.10)(0.12,-0.16){5}{\textcolor{red}{\line(0,-1){0.16}}}
\multiput(31.31,18.33)(0.13,-0.15){5}{\textcolor{red}{\line(0,-1){0.15}}}
\multiput(31.95,17.57)(0.13,-0.15){5}{\textcolor{red}{\line(0,-1){0.15}}}
\multiput(32.60,16.82)(0.11,-0.12){6}{\textcolor{red}{\line(0,-1){0.12}}}
\multiput(33.27,16.08)(0.11,-0.12){6}{\textcolor{red}{\line(0,-1){0.12}}}
\multiput(33.95,15.35)(0.12,-0.12){6}{\textcolor{red}{\line(0,-1){0.12}}}
\multiput(34.64,14.64)(0.12,-0.12){6}{\textcolor{red}{\line(1,0){0.12}}}
\multiput(35.35,13.95)(0.12,-0.11){6}{\textcolor{red}{\line(1,0){0.12}}}
\multiput(36.08,13.27)(0.12,-0.11){6}{\textcolor{red}{\line(1,0){0.12}}}
\multiput(36.82,12.60)(0.15,-0.13){5}{\textcolor{red}{\line(1,0){0.15}}}
\multiput(37.57,11.95)(0.15,-0.13){5}{\textcolor{red}{\line(1,0){0.15}}}
\multiput(38.33,11.31)(0.16,-0.12){5}{\textcolor{red}{\line(1,0){0.16}}}
\multiput(39.10,10.69)(0.16,-0.12){5}{\textcolor{red}{\line(1,0){0.16}}}
\multiput(39.89,10.08)(0.16,-0.12){5}{\textcolor{red}{\line(1,0){0.16}}}
\multiput(40.69,9.49)(0.16,-0.11){5}{\textcolor{red}{\line(1,0){0.16}}}
\multiput(41.50,8.92)(0.16,-0.11){5}{\textcolor{red}{\line(1,0){0.16}}}
\multiput(42.32,8.36)(0.17,-0.11){5}{\textcolor{red}{\line(1,0){0.17}}}
\multiput(43.16,7.82)(0.21,-0.13){4}{\textcolor{red}{\line(1,0){0.21}}}
\multiput(44.00,7.29)(0.21,-0.13){4}{\textcolor{red}{\line(1,0){0.21}}}
\multiput(44.86,6.78)(0.22,-0.12){4}{\textcolor{red}{\line(1,0){0.22}}}
\multiput(45.72,6.29)(0.22,-0.12){4}{\textcolor{red}{\line(1,0){0.22}}}
\multiput(46.59,5.82)(0.22,-0.11){4}{\textcolor{red}{\line(1,0){0.22}}}
\multiput(47.48,5.36)(0.22,-0.11){4}{\textcolor{red}{\line(1,0){0.22}}}
\multiput(48.37,4.92)(0.23,-0.11){4}{\textcolor{red}{\line(1,0){0.23}}}
\multiput(49.27,4.50)(0.30,-0.13){3}{\textcolor{red}{\line(1,0){0.30}}}
\multiput(50.18,4.10)(0.31,-0.13){3}{\textcolor{red}{\line(1,0){0.31}}}
\multiput(51.10,3.71)(0.31,-0.12){3}{\textcolor{red}{\line(1,0){0.31}}}
\multiput(52.02,3.34)(0.31,-0.12){3}{\textcolor{red}{\line(1,0){0.31}}}
\multiput(52.95,3.00)(0.31,-0.11){3}{\textcolor{red}{\line(1,0){0.31}}}
\multiput(53.89,2.67)(0.31,-0.10){3}{\textcolor{red}{\line(1,0){0.31}}}
\multiput(54.83,2.36)(0.48,-0.15){2}{\textcolor{red}{\line(1,0){0.48}}}
\multiput(55.78,2.06)(0.48,-0.14){2}{\textcolor{red}{\line(1,0){0.48}}}
\multiput(56.74,1.79)(0.48,-0.13){2}{\textcolor{red}{\line(1,0){0.48}}}
\multiput(57.70,1.54)(0.48,-0.12){2}{\textcolor{red}{\line(1,0){0.48}}}
\multiput(58.67,1.30)(0.49,-0.11){2}{\textcolor{red}{\line(1,0){0.49}}}
\multiput(59.64,1.09)(0.49,-0.10){2}{\textcolor{red}{\line(1,0){0.49}}}
\multiput(60.61,0.89)(0.98,-0.18){1}{\textcolor{red}{\line(1,0){0.98}}}
\multiput(61.59,0.71)(0.98,-0.16){1}{\textcolor{red}{\line(1,0){0.98}}}
\multiput(62.57,0.55)(0.98,-0.14){1}{\textcolor{red}{\line(1,0){0.98}}}
\multiput(63.56,0.42)(0.99,-0.12){1}{\textcolor{red}{\line(1,0){0.99}}}
\multiput(64.54,0.30)(0.99,-0.10){1}{\textcolor{red}{\line(1,0){0.99}}}
\multiput(65.53,0.20)(0.99,-0.08){1}{\textcolor{red}{\line(1,0){0.99}}}
\multiput(66.52,0.12)(0.99,-0.06){1}{\textcolor{red}{\line(1,0){0.99}}}
\multiput(67.52,0.06)(0.99,-0.04){1}{\textcolor{red}{\line(1,0){0.99}}}
\multiput(68.51,0.02)(0.99,-0.02){1}{\textcolor{red}{\line(1,0){0.99}}}
\put(69.0,0.00){\textcolor{red}{\line(1,0){0.99}}}
\linethickness{0.25mm}
\put(80.00,0.00){\textcolor{blue}{\line(0,1){50.00}}}
\put(80.00,50.00){\textcolor{blue}{\vector(0,1){0.}}}
\put(60.00,50.00){\textcolor{blue}{\line(1,0){20.00}}}
\put(60.00,50.00){\textcolor{blue}{\vector(-1,0){0.}}}
\put(60.00,50.00){\textcolor{blue}{\line(0,1){45.50}}}
\put(60.00,95.50){\textcolor{blue}{\vector(0,1){0.}}}
\linethickness{0.25mm}
\multiput(59.2,95.50)(0.30,0.13){3}{\textcolor{blue}{\line(1,0){0.30}}}
\multiput(58.37,95.08)(0.23,0.11){4}{\textcolor{blue}{\line(1,0){0.23}}}
\multiput(57.48,94.64)(0.22,0.11){4}{\textcolor{blue}{\line(1,0){0.22}}}
\multiput(56.59,94.18)(0.22,0.11){4}{\textcolor{blue}{\line(1,0){0.22}}}
\multiput(55.72,93.71)(0.22,0.12){4}{\textcolor{blue}{\line(1,0){0.22}}}
\multiput(54.86,93.22)(0.22,0.12){4}{\textcolor{blue}{\line(1,0){0.22}}}
\multiput(54.00,92.71)(0.21,0.13){4}{\textcolor{blue}{\line(1,0){0.21}}}
\multiput(53.16,92.18)(0.21,0.13){4}{\textcolor{blue}{\line(1,0){0.21}}}
\multiput(52.32,91.64)(0.17,0.11){5}{\textcolor{blue}{\line(1,0){0.17}}}
\multiput(51.50,91.08)(0.16,0.11){5}{\textcolor{blue}{\line(1,0){0.16}}}
\multiput(50.69,90.51)(0.16,0.11){5}{\textcolor{blue}{\line(1,0){0.16}}}
\multiput(49.89,89.92)(0.16,0.12){5}{\textcolor{blue}{\line(1,0){0.16}}}
\multiput(49.10,89.31)(0.16,0.12){5}{\textcolor{blue}{\line(1,0){0.16}}}
\multiput(48.33,88.69)(0.16,0.12){5}{\textcolor{blue}{\line(1,0){0.16}}}
\multiput(47.57,88.05)(0.15,0.13){5}{\textcolor{blue}{\line(1,0){0.15}}}
\multiput(46.82,87.40)(0.15,0.13){5}{\textcolor{blue}{\line(1,0){0.15}}}
\multiput(46.08,86.73)(0.12,0.11){6}{\textcolor{blue}{\line(1,0){0.12}}}
\multiput(45.35,86.05)(0.12,0.11){6}{\textcolor{blue}{\line(1,0){0.12}}}
\multiput(44.64,85.36)(0.12,0.12){6}{\textcolor{blue}{\line(1,0){0.12}}}
\multiput(43.95,84.65)(0.12,0.12){6}{\textcolor{blue}{\line(0,1){0.12}}}
\multiput(43.27,83.92)(0.11,0.12){6}{\textcolor{blue}{\line(0,1){0.12}}}
\multiput(42.60,83.18)(0.11,0.12){6}{\textcolor{blue}{\line(0,1){0.12}}}
\multiput(41.95,82.43)(0.13,0.15){5}{\textcolor{blue}{\line(0,1){0.15}}}
\multiput(41.31,81.67)(0.13,0.15){5}{\textcolor{blue}{\line(0,1){0.15}}}
\multiput(40.69,80.90)(0.12,0.16){5}{\textcolor{blue}{\line(0,1){0.16}}}
\multiput(40.08,80.11)(0.12,0.16){5}{\textcolor{blue}{\line(0,1){0.16}}}
\multiput(39.49,79.31)(0.12,0.16){5}{\textcolor{blue}{\line(0,1){0.16}}}
\multiput(38.92,78.50)(0.11,0.16){5}{\textcolor{blue}{\line(0,1){0.16}}}
\multiput(38.36,77.68)(0.11,0.16){5}{\textcolor{blue}{\line(0,1){0.16}}}
\multiput(37.82,76.84)(0.11,0.17){5}{\textcolor{blue}{\line(0,1){0.17}}}
\multiput(37.29,76.00)(0.13,0.21){4}{\textcolor{blue}{\line(0,1){0.21}}}
\multiput(36.78,75.14)(0.13,0.21){4}{\textcolor{blue}{\line(0,1){0.21}}}
\multiput(36.29,74.28)(0.12,0.22){4}{\textcolor{blue}{\line(0,1){0.22}}}
\multiput(35.82,73.41)(0.12,0.22){4}{\textcolor{blue}{\line(0,1){0.22}}}
\multiput(35.36,72.52)(0.11,0.22){4}{\textcolor{blue}{\line(0,1){0.22}}}
\multiput(34.92,71.63)(0.11,0.22){4}{\textcolor{blue}{\line(0,1){0.22}}}
\multiput(34.50,70.73)(0.11,0.23){4}{\textcolor{blue}{\line(0,1){0.23}}}
\multiput(34.10,69.82)(0.13,0.30){3}{\textcolor{blue}{\line(0,1){0.30}}}
\multiput(33.71,68.90)(0.13,0.31){3}{\textcolor{blue}{\line(0,1){0.31}}}
\multiput(33.34,67.98)(0.12,0.31){3}{\textcolor{blue}{\line(0,1){0.31}}}
\multiput(33.00,67.05)(0.12,0.31){3}{\textcolor{blue}{\line(0,1){0.31}}}
\multiput(32.67,66.11)(0.11,0.31){3}{\textcolor{blue}{\line(0,1){0.31}}}
\multiput(32.36,65.17)(0.10,0.31){3}{\textcolor{blue}{\line(0,1){0.31}}}
\multiput(32.06,64.22)(0.15,0.48){2}{\textcolor{blue}{\line(0,1){0.48}}}
\multiput(31.79,63.26)(0.14,0.48){2}{\textcolor{blue}{\line(0,1){0.48}}}
\multiput(31.54,62.30)(0.13,0.48){2}{\textcolor{blue}{\line(0,1){0.48}}}
\multiput(31.30,61.33)(0.12,0.48){2}{\textcolor{blue}{\line(0,1){0.48}}}
\multiput(31.09,60.36)(0.11,0.49){2}{\textcolor{blue}{\line(0,1){0.49}}}
\multiput(30.89,59.39)(0.10,0.49){2}{\textcolor{blue}{\line(0,1){0.49}}}
\multiput(30.71,58.41)(0.18,0.98){1}{\textcolor{blue}{\line(0,1){0.98}}}
\multiput(30.55,57.43)(0.16,0.98){1}{\textcolor{blue}{\line(0,1){0.98}}}
\multiput(30.42,56.44)(0.14,0.98){1}{\textcolor{blue}{\line(0,1){0.98}}}
\multiput(30.30,55.46)(0.12,0.99){1}{\textcolor{blue}{\line(0,1){0.99}}}
\multiput(30.20,54.47)(0.10,0.99){1}{\textcolor{blue}{\line(0,1){0.99}}}
\multiput(30.12,53.48)(0.08,0.99){1}{\textcolor{blue}{\line(0,1){0.99}}}
\multiput(30.06,52.48)(0.06,0.99){1}{\textcolor{blue}{\line(0,1){0.99}}}
\multiput(30.02,51.49)(0.04,0.99){1}{\textcolor{blue}{\line(0,1){0.99}}}
\multiput(30.00,50.50)(0.02,0.99){1}{\textcolor{blue}{\line(0,1){0.99}}}
\put(30.00,50.00){\textcolor{blue}{\vector(0,-1){0.12}}}
\put(30.00,49.50){\textcolor{blue}{\line(0,1){0.99}}}
\multiput(30.00,49.50)(0.02,-0.99){1}{\textcolor{blue}{\line(0,-1){0.99}}}
\multiput(30.02,48.51)(0.04,-0.99){1}{\textcolor{blue}{\line(0,-1){0.99}}}
\multiput(30.06,47.52)(0.06,-0.99){1}{\textcolor{blue}{\line(0,-1){0.99}}}
\multiput(30.12,46.52)(0.08,-0.99){1}{\textcolor{blue}{\line(0,-1){0.99}}}
\multiput(30.20,45.53)(0.10,-0.99){1}{\textcolor{blue}{\line(0,-1){0.99}}}
\multiput(30.30,44.54)(0.12,-0.99){1}{\textcolor{blue}{\line(0,-1){0.99}}}
\multiput(30.42,43.56)(0.14,-0.98){1}{\textcolor{blue}{\line(0,-1){0.98}}}
\multiput(30.55,42.57)(0.16,-0.98){1}{\textcolor{blue}{\line(0,-1){0.98}}}
\multiput(30.71,41.59)(0.18,-0.98){1}{\textcolor{blue}{\line(0,-1){0.98}}}
\multiput(30.89,40.61)(0.10,-0.49){2}{\textcolor{blue}{\line(0,-1){0.49}}}
\multiput(31.09,39.64)(0.11,-0.49){2}{\textcolor{blue}{\line(0,-1){0.49}}}
\multiput(31.30,38.67)(0.12,-0.48){2}{\textcolor{blue}{\line(0,-1){0.48}}}
\multiput(31.54,37.70)(0.13,-0.48){2}{\textcolor{blue}{\line(0,-1){0.48}}}
\multiput(31.79,36.74)(0.14,-0.48){2}{\textcolor{blue}{\line(0,-1){0.48}}}
\multiput(32.06,35.78)(0.15,-0.48){2}{\textcolor{blue}{\line(0,-1){0.48}}}
\multiput(32.36,34.83)(0.10,-0.31){3}{\textcolor{blue}{\line(0,-1){0.31}}}
\multiput(32.67,33.89)(0.11,-0.31){3}{\textcolor{blue}{\line(0,-1){0.31}}}
\multiput(33.00,32.95)(0.12,-0.31){3}{\textcolor{blue}{\line(0,-1){0.31}}}
\multiput(33.34,32.02)(0.12,-0.31){3}{\textcolor{blue}{\line(0,-1){0.31}}}
\multiput(33.71,31.10)(0.13,-0.31){3}{\textcolor{blue}{\line(0,-1){0.31}}}
\multiput(34.10,30.18)(0.13,-0.30){3}{\textcolor{blue}{\line(0,-1){0.30}}}
\multiput(34.50,29.27)(0.11,-0.23){4}{\textcolor{blue}{\line(0,-1){0.23}}}
\multiput(34.92,28.37)(0.11,-0.22){4}{\textcolor{blue}{\line(0,-1){0.22}}}
\multiput(35.36,27.48)(0.11,-0.22){4}{\textcolor{blue}{\line(0,-1){0.22}}}
\multiput(35.82,26.59)(0.12,-0.22){4}{\textcolor{blue}{\line(0,-1){0.22}}}
\multiput(36.29,25.72)(0.12,-0.22){4}{\textcolor{blue}{\line(0,-1){0.22}}}
\multiput(36.78,24.86)(0.13,-0.21){4}{\textcolor{blue}{\line(0,-1){0.21}}}
\multiput(37.29,24.00)(0.13,-0.21){4}{\textcolor{blue}{\line(0,-1){0.21}}}
\multiput(37.82,23.16)(0.11,-0.17){5}{\textcolor{blue}{\line(0,-1){0.17}}}
\multiput(38.36,22.32)(0.11,-0.16){5}{\textcolor{blue}{\line(0,-1){0.16}}}
\multiput(38.92,21.50)(0.11,-0.16){5}{\textcolor{blue}{\line(0,-1){0.16}}}
\multiput(39.49,20.69)(0.12,-0.16){5}{\textcolor{blue}{\line(0,-1){0.16}}}
\multiput(40.08,19.89)(0.12,-0.16){5}{\textcolor{blue}{\line(0,-1){0.16}}}
\multiput(40.69,19.10)(0.12,-0.16){5}{\textcolor{blue}{\line(0,-1){0.16}}}
\multiput(41.31,18.33)(0.13,-0.15){5}{\textcolor{blue}{\line(0,-1){0.15}}}
\multiput(41.95,17.57)(0.13,-0.15){5}{\textcolor{blue}{\line(0,-1){0.15}}}
\multiput(42.60,16.82)(0.11,-0.12){6}{\textcolor{blue}{\line(0,-1){0.12}}}
\multiput(43.27,16.08)(0.11,-0.12){6}{\textcolor{blue}{\line(0,-1){0.12}}}
\multiput(43.95,15.35)(0.12,-0.12){6}{\textcolor{blue}{\line(0,-1){0.12}}}
\multiput(44.64,14.64)(0.12,-0.12){6}{\textcolor{blue}{\line(1,0){0.12}}}
\multiput(45.35,13.95)(0.12,-0.11){6}{\textcolor{blue}{\line(1,0){0.12}}}
\multiput(46.08,13.27)(0.12,-0.11){6}{\textcolor{blue}{\line(1,0){0.12}}}
\multiput(46.82,12.60)(0.15,-0.13){5}{\textcolor{blue}{\line(1,0){0.15}}}
\multiput(47.57,11.95)(0.15,-0.13){5}{\textcolor{blue}{\line(1,0){0.15}}}
\multiput(48.33,11.31)(0.16,-0.12){5}{\textcolor{blue}{\line(1,0){0.16}}}
\multiput(49.10,10.69)(0.16,-0.12){5}{\textcolor{blue}{\line(1,0){0.16}}}
\multiput(49.89,10.08)(0.16,-0.12){5}{\textcolor{blue}{\line(1,0){0.16}}}
\multiput(50.69,9.49)(0.16,-0.11){5}{\textcolor{blue}{\line(1,0){0.16}}}
\multiput(51.50,8.92)(0.16,-0.11){5}{\textcolor{blue}{\line(1,0){0.16}}}
\multiput(52.32,8.36)(0.17,-0.11){5}{\textcolor{blue}{\line(1,0){0.17}}}
\multiput(53.16,7.82)(0.21,-0.13){4}{\textcolor{blue}{\line(1,0){0.21}}}
\multiput(54.00,7.29)(0.21,-0.13){4}{\textcolor{blue}{\line(1,0){0.21}}}
\multiput(54.86,6.78)(0.22,-0.12){4}{\textcolor{blue}{\line(1,0){0.22}}}
\multiput(55.72,6.29)(0.22,-0.12){4}{\textcolor{blue}{\line(1,0){0.22}}}
\multiput(56.59,5.82)(0.22,-0.11){4}{\textcolor{blue}{\line(1,0){0.22}}}
\multiput(57.48,5.36)(0.22,-0.11){4}{\textcolor{blue}{\line(1,0){0.22}}}
\multiput(58.37,4.92)(0.23,-0.11){4}{\textcolor{blue}{\line(1,0){0.23}}}
\multiput(59.27,4.50)(0.30,-0.13){3}{\textcolor{blue}{\line(1,0){0.30}}}
\multiput(60.18,4.10)(0.31,-0.13){3}{\textcolor{blue}{\line(1,0){0.31}}}
\multiput(61.10,3.71)(0.31,-0.12){3}{\textcolor{blue}{\line(1,0){0.31}}}
\multiput(62.02,3.34)(0.31,-0.12){3}{\textcolor{blue}{\line(1,0){0.31}}}
\multiput(62.95,3.00)(0.31,-0.11){3}{\textcolor{blue}{\line(1,0){0.31}}}
\multiput(63.89,2.67)(0.31,-0.10){3}{\textcolor{blue}{\line(1,0){0.31}}}
\multiput(64.83,2.36)(0.48,-0.15){2}{\textcolor{blue}{\line(1,0){0.48}}}
\multiput(65.78,2.06)(0.48,-0.14){2}{\textcolor{blue}{\line(1,0){0.48}}}
\multiput(66.74,1.79)(0.48,-0.13){2}{\textcolor{blue}{\line(1,0){0.48}}}
\multiput(67.70,1.54)(0.48,-0.12){2}{\textcolor{blue}{\line(1,0){0.48}}}
\multiput(68.67,1.30)(0.49,-0.11){2}{\textcolor{blue}{\line(1,0){0.49}}}
\multiput(69.64,1.09)(0.49,-0.10){2}{\textcolor{blue}{\line(1,0){0.49}}}
\multiput(70.61,0.89)(0.98,-0.18){1}{\textcolor{blue}{\line(1,0){0.98}}}
\multiput(71.59,0.71)(0.98,-0.16){1}{\textcolor{blue}{\line(1,0){0.98}}}
\multiput(72.57,0.55)(0.98,-0.14){1}{\textcolor{blue}{\line(1,0){0.98}}}
\multiput(73.56,0.42)(0.99,-0.12){1}{\textcolor{blue}{\line(1,0){0.99}}}
\multiput(74.54,0.30)(0.99,-0.10){1}{\textcolor{blue}{\line(1,0){0.99}}}
\multiput(75.53,0.20)(0.99,-0.08){1}{\textcolor{blue}{\line(1,0){0.99}}}
\multiput(76.52,0.12)(0.99,-0.06){1}{\textcolor{blue}{\line(1,0){0.99}}}
\multiput(77.52,0.06)(0.99,-0.04){1}{\textcolor{blue}{\line(1,0){0.99}}}
\multiput(78.51,0.02)(0.99,-0.02){1}{\textcolor{blue}{\line(1,0){0.99}}}
\put(79.0,0.00){\textcolor{blue}{\line(1,0){0.99}}}
\end{picture}
\caption{Integration contours $\Gamma$ and $\Gamma'$ in the complex plane
$z=x+iy$, which are used to evaluate the Laplace antitransform of 
$\widetilde{\mathcal{W}}_N(z)$. 
The saddle point $x_0$ and the poles of $\widetilde{\mathcal{W}}_N(z)$ 
are indicated by open and filled circles, respectively.}
\label{contours}
\end{figure}

The fact that $y_{0}=0$ also implies
\begin{eqnarray}
\partial_z^2 \varphi(z_{0}) =
\sum_{V \in \mathscr{V}}\frac{N_{V}}{N}\frac{1}{(x_{0}+V)^{2}},
\end{eqnarray}
so that we have 
\begin{eqnarray}
\Arg \partial_z^2 \varphi(z_{0}) =0, 
\end{eqnarray}
\textit{i.e.} $\alpha=\pi/2$.
The integration contour $\Gamma$, therefore, has to be deformed into a new
one $\Gamma'$ parallel to the real axis.   
At a first sight, this kind of deformation is incompatible with the condition
that $\Gamma'$ strictly contains all the poles $z^{}_{V}$ 
located on the real axis.
However, the largest solution of Eq.~(\ref{X})
is bounded from below by $x_{0} > -V_{\mathrm{min}} +N_{V_{\mathrm{min}}}/t$.
This means that for $t\to\infty$ with 
$N_{V_{\mathrm{min}}} \sim N \sim t$, the distance between the 
saddle point and the closest pole at $z=-V_{\mathrm{min}}$ is finite.
Therefore, we can take $\Gamma'$ parallel to the real axis only in a finite
neighborhood of the saddle point as shown in Fig.~\ref{contours}.
In fact, the contribution to the integral due to this neighborhood of $z_0$
is exponentially leading in $N$ with respect to the rest of the contour.

In conclusion, for $t \to \infty$ with $N\sim t$, the integration on 
the contour $\Gamma'$ chosen as described above becomes asymptotically 
exact and, in this limit, we have the following exact expression 
for the weights
\numparts
\begin{eqnarray}
\mathcal{W}_{N}(t) =
\frac{e^{x_{0}t-\sum_{V\in\mathscr{V}} N_{V} \log[(x_{0}+V)/\epsilon]}}
{\sqrt{ 2\pi \sum_{V\in\mathscr{V}}\frac{\epsilon^2 N_{V}}{(x_{0}+V)^{2}} } },
\label{WSADDLE1}
\\ 
\sum_{V \in \mathscr{V}}\frac{N_{V}}{x_{0}+V}=t, 
\qquad x_{0}>-V_{\mathrm{min}}.
\label{WSADDLE2}
\end{eqnarray}
\endnumparts
This expression is semi-analytic in the sense that the simple 
Eq.~(\ref{WSADDLE2}), 
which provides $x_0$ to be inserted into Eq. (\ref{WSADDLE1}),
must be solved, in general, numerically or recursively.

For $\hat{V}\equiv 0$, we can verify that the above expression
for the weights reproduces the exact result 
$\mathcal{W}_{N}^{(0)}(t)=\epsilon^N t^{N}/N!$. 
In this case $\mathscr{V}=\{0\}$ and the solution of Eq.~(\ref{WSADDLE2}) is
\begin{eqnarray}
\label{X0}
x_{0} = \frac{N+1}{t}.
\end{eqnarray}
Inserting this value into Eq.~(\ref{WSADDLE1}), we have
\begin{eqnarray}
\label{W0}
\mathcal{W}_{N}^{(0)}(t)&=&
\frac{\exp\left[N+1-(N+1)\log\left[(N+1)/\epsilon t \right]\right]}
{\sqrt{ 2 \pi \epsilon^2 t^2 /(N+1) } }\nonumber \\
&=& \frac{1}{\sqrt{2\pi}}
\frac{\exp(N+1)}{(N+1)^{N+1/2}}~\epsilon^Nt^{N}. 
\end{eqnarray}
By recalling Stirling's formula 
\begin{eqnarray}
\label{STIRLING}
N!\simeq \sqrt{2\pi} N^{N+1/2} e^{-N},
\end{eqnarray}
which derives from a saddle-point evaluation of the Gamma function 
as well, we see that Eq.~(\ref{W0}) is just the Stirling approximation 
to $\epsilon^N t^{N}/N!$.

\subsection{Canonical averages via large deviation analysis}
\label{largedeviation}

To evaluate the canonical averages it is useful to introduce
the frequencies, $\nu^{}_V=N_V/N$, $V \in \mathscr{V}$, 
and $\nu^{}_T=N_T/N$, $T \in \mathscr{T}$, 
which for $N$ large become continuously distributed in the range
$[0,1]$ with the constraints 
\begin{eqnarray}
\label{CONSTRAINTS}
\sum_{V \in \mathscr{V}} \nu^{}_V = \sum_{T\in \mathscr{T}} \nu^{}_T =1.
\end{eqnarray}
Note that, for $N$ large, we do not distinguish the different normalizations,
$N+1$ and $N$, of $N_V$ and $N_T$, respectively.
When possible, we will use a compact notation in terms of the vectors
$\bm{\mu}$ and $\bm{\nu}$, which have
$m=m_{\mathscr{V}}+m_{\mathscr{T}}$ components 
indicated by a Greek index 
$\alpha \in \mathscr{H} = \mathscr{V} \cup \mathscr{T}$
and are defined as
$\bm{\mu}^\mathrm{T}=(\ldots N_V \ldots ; \ldots N_T \ldots)$ and
$\bm{\nu}^\mathrm{T} = (\ldots \nu^{}_V \ldots; \ldots \nu^{}_T \ldots)$.
We have
\begin{eqnarray}
\label{MU}
\bm{\mu}=N\bm{\nu}.
\end{eqnarray}
For later use, we also define 
$\bm{u}^\mathrm{T}= (\ldots -\log[(x_{0}+V)/\epsilon] \ldots;
\ldots \log T \ldots)$,
$\bm{v}^\mathrm{T} = (\ldots (x_{0}+V)^{-1} \ldots; \ldots 0 \ldots)$
and 
$\bm{w}^\mathrm{T} = (\ldots (x_{0}+V)^{-2} \ldots; \ldots 0 \ldots)$.
Note that the vectors $\bm{u}$, $\bm{v}$ and $\bm{w}$ 
depend on $\bm{\nu}$ through $x_{0}=x_{0}(\bm{\nu})$ and
$\bm{v}=-\partial_{x_{0}}\bm{u}$, $\bm{w}=-\partial_{x_{0}}\bm{v}$.
Finally, we will take advantage of a scalar product notation.
For instance, we rewrite the saddle-point Eq.~(\ref{WSADDLE2}) as 
$(\bm{\nu},\bm{v})=t/N$. 

By using the result given by Eqs. (\ref{WSADDLE1}) and (\ref{WSADDLE2}), 
we express the r.h.s. of Eq.~(\ref{AVERAGE}) in the following explicit form
\begin{eqnarray}
\label{AVERAGE1}
\media{ \mathcal{W}_{N}(t) \! \prod_{T\in \mathscr{T}} T^{N_T} }  
= \sum_{\bm{\mu}}
\mathcal{P}_N(\bm{\mu}) 
\frac{e^{x_{0}t + \left( \bm{\mu},\bm{u} \right) }}
{ \sqrt {2\pi \epsilon^2 (\bm{\mu},\bm{w}) }}.
\end{eqnarray}
The probability $\mathcal{P}_N(\bm{\mu})$ is
given by the fraction of trajectories 
branching from the initial configuration $\bm{n}_0$ and having
after $N$ jumps multiplicities $\bm{\mu}$.
According to Poisson's summation formula
\begin{eqnarray}
\label{POISSON}
\sum_{\bm{\mu}}f(\bm{\mu})=
\sum_{\bm{k} \in {\mathbb{Z}^m_{}}}\int 
d\bm{\mu}e^{i(\bm{k},\bm{\mu})}f(\bm{\mu}),
\end{eqnarray}
the sum over $\bm{\mu}$ in Eq.~(\ref{AVERAGE1}) can be transformed
into a series for $\bm{k} \in {\mathbb{Z}^m_{}}$ 
of integrals over the same variable in the presence 
of the oscillating factors $\exp[i(\bm{k},\bm{\mu})]$.
As we will check at the end of the calculation, in the limit of $t$ large
all the terms $\bm{k}\neq \bm{0}$ of this series are exponentially damped
with respect to the term $\bm{k}=\bm{0}$.
In this limit, therefore, we will not distinguish the sum over $\bm{\mu}$ 
in Eq.~(\ref{AVERAGE1}) with the corresponding integral. 

In \cite{OP1} we have evaluated Eq.~(\ref{AVERAGE1}) for $t$ large
by using a central limit theorem for Markov chains. 
Although this theorem applies rigorously to each rescaled sum 
$N_{\alpha}/\sqrt{N} = \nu_\alpha\sqrt{N}$, 
$\alpha \in \mathscr{H}$,
it provides an approximation when, as in Eq.~(\ref{AVERAGE1}),
variables like $\nu_{\alpha}N$ are involved.
As anticipated in~\cite{OP1},
the integrand  in Eq.~(\ref{AVERAGE1}) is sensitive 
to the large deviations of $\mathcal{P}_N(\bm{\mu})$ 
from the central limit behavior
and to obtain more accurate estimates we need 
to consider a development in terms of the associated cumulants
(connected correlation functions) \cite{SHY}.

Before proceeding with this analysis, we observe that 
the constraints (\ref{CONSTRAINTS}) give important  
summation rules for the cumulants.
Let us indicate with 
$\mediac{\nu_{\alpha_1}\dots\nu_{\alpha_k}}$,
$ \alpha_1,\dots, \alpha_{k} \in\mathscr{H}$, 
a cumulant of order $k$.
In Appendix \ref{csr-proof}, we demonstrate that 
\begin{eqnarray}
\label{c.c.f.1}
\sum_{\alpha \in \mathscr{V}}\mediac{\nu_\alpha}=
\sum_{\alpha \in \mathscr{T}}\mediac{\nu_\alpha}=1,
\end{eqnarray}
for $k=1$, whereas for $k>1$
\begin{eqnarray}
\label{c.c.f.}
\sum_{\alpha_k \in \mathscr{V}}
\mediac{\nu_{\alpha_1}\ldots\nu_{\alpha_k}}=
\sum_{\alpha_k \in \mathscr{T}}
\mediac{\nu_{\alpha_1}\ldots\nu_{\alpha_k}} =0.
\end{eqnarray}
These rules provide
a basic test for the statistical measurement of the cumulants themselves.
A sampling that aims at measuring the cumulants with
a given accuracy, will have to satisfy Eqs.~(\ref{c.c.f.1}) and (\ref{c.c.f.})
with the same accuracy.  

As customary in the framework of a large deviation analysis,
we are interested to get information about the density
$\mathcal{P}_N(N\bm{\nu})$ in the limit of $N$ large and $\bm{\nu}$ finite. 
By introducing the Fourier anti-transformation
\begin{eqnarray}
\label{FOURIER}
\mathcal{P}_N(\bm{\mu})= \frac{1}{(2\pi)^{m}}
\int d\bm{q}
e^{\log \left[ {\widetilde{\mathcal{P}}}_N(\bm{q}) \right]
-i (\bm{q},\bm{\mu})},
\end{eqnarray}
${\widetilde{\mathcal{P}}}_N(\bm{q})$ being the Fourier transform of
$\mathcal{P}_N(\bm{\mu})$,
Eq.~(\ref{AVERAGE1}) becomes 
\begin{eqnarray}
\label{AVERAGE2}
\media{ \mathcal{W}_{N}(t) \! \prod_{T\in \mathscr{T}} T^{N_T} } 
=
\left(\frac{N}{2\pi}\right)^{m} 
\int d\bm{\nu} \int d\bm{q}
e^{N\phi(\bm{\nu},\bm{q})}R(\bm{\nu}),
\end{eqnarray}
where $\phi(\bm{\nu},\bm{q})$ takes into account the exponential
behavior of the integrand, whereas $R(\bm{\nu})$ is a smooth function.
For brevity, we omit the parametric dependence of $\phi$ and $R$ 
on $t$ and $N$. 
Explicitly, we have 
\begin{eqnarray}
\phi(\bm{\nu},\bm{q}) &=&
x_{0}\frac{t}{N} + \left( \bm{\nu},\bm{u} \right)
-i\left( \bm{\nu},\bm{q} \right) +
\frac{\log {\widetilde{\mathcal{P}}}_N(\bm{q}) }{N} 
\quad \\
R(\bm{\nu}) &=& 
\frac{1} { \sqrt {2\pi N \epsilon^2 (\bm{\nu},\bm{w}) }}.
\end{eqnarray}

As well known, the cumulants are related to 
$\log{\widetilde{\mathcal{P}}}_N(\bm{q})$ through the relation  
\begin{eqnarray}
\label{LOGP}
\log {\widetilde{\mathcal{P}}}_N(\bm{q}) &=&
\log\media{e^{i\left( \bm{\mu},\bm{q} \right)}}
\nonumber \\ &=& 
\sum_{k=1}^{\infty} \frac{1}{k!}
\mediac{\left( \bm{\mu},i\bm{q} \right)^k},
\nonumber \\ &=&
\sum_{k=1}^{\infty} \frac{N^k}{k!}
\mediac{\left( \bm{\nu},i\bm{q} \right)^k}.
\end{eqnarray}
Note that for any given $N$, due to the inequalities 
$\media{\mu_{\alpha_1}\dots\mu_{\alpha_k}} \leq N^k$,
valid for any $k$, 
and due to the asymmetry 
$\mathcal{P}_N(\bm{\mu}) \neq \mathcal{P}_N(-\bm{\mu})$,
the series in Eq.~(\ref{LOGP}) converge 
for every $\bm{q} \in \mathbb{C}^m$  (see, for example~\cite{SHY}).

We evaluate the integrals in Eq.~(\ref{AVERAGE2}) 
by a saddle-point calculation (actually Laplace's method) 
with respect to the variable 
$(\bm{\nu},\bm{q}) \in \mathbb{R}^m\times \mathbb{R}^m$.
The derivatives of $\phi(\bm{\nu},\bm{q})$ with respect to 
$q_{\alpha}$ and $\nu_{\alpha}$, 
$\alpha \in \mathscr{H}$, are
\begin{eqnarray*}
\partial_{q_{\alpha}} \phi(\bm{\nu},\bm{q}) &=& 
i\sum_{k=1}^{\infty} \frac{N^{k-1}}{(k-1)!}
\mediac{\left( \bm{\nu},i\bm{q} \right)^{k-1}\nu_{\alpha}}
 -i \nu_{\alpha}
\\ 
\partial_{\nu_{\alpha}}\phi(\bm{\nu},\bm{q}) &=&
-iq_{\alpha}+u_{\alpha},
\end{eqnarray*}
where we have used $\left( \bm{\nu},\bm{v} \right)=t/N$.
Therefore, the stationarity condition for $\phi(\bm{\nu},\bm{q})$ implies
\begin{eqnarray*}
\nu_{\alpha} &=& 
\sum_{k=1}^{\infty} \frac{N^{k-1}}{(k-1)!}
\mediac{\left( \bm{\nu},i\bm{q} \right)^{k-1}\nu_{\alpha}}
\\ 
iq_{\alpha} &=& u_{\alpha},
\end{eqnarray*}
which, in turn, reduces to the following equation
for the saddle-point frequencies $\bm{\nu}^\mathrm{sp}$ 
\begin{eqnarray}
\label{SPEQ}
\nu_{\alpha}^\mathrm{sp} = 
\sum_{k=1}^{\infty} \frac{N^{k-1}}{(k-1)!}
\mediac{\left( \bm{\nu},\bm{u}(x_{0}(\bm{\nu}^\mathrm{sp})) 
\right)^{k-1}\nu_{\alpha}} . 
\end{eqnarray}
Hereafter, we will add the superscript $^{\mathrm{sp}}$ to a function 
of $\bm{\nu}$ to indicate the value of this function for
$\bm{\nu}=\bm{\nu}^\mathrm{sp}$. 
By approximating the function $\phi(\bm{\nu},\bm{q})$ with its 
second order Taylor expansion around the saddle point 
$(\bm{\nu}^\mathrm{sp},\bm{q}^\mathrm{sp})$
and performing the resulting Gaussian integral with respect to the variable 
$(\bm{\nu}-\bm{\nu}^\mathrm{sp},\bm{q}-\bm{q}^\mathrm{sp})$,
Eq.~(\ref{AVERAGE2}) becomes
\begin{eqnarray}
\label{AVERAGE3}
\media{\mathcal{W}_{N}(t) \! \prod_{T\in \mathscr{T}} T^{N_T} } = 
C_{N}\exp{\left[\psi(N)\right]},
\end{eqnarray}
where
\begin{eqnarray}
\label{PSI}
\psi(N) &=& 
N\phi(\bm{{\nu}}^\mathrm{sp},\bm{q}^\mathrm{sp}) \nonumber \\
&=& x_{0}^{\mathrm{sp}}t+ 
N\sum_{k=1}^{\infty} \frac{N^{k-1}}{k!}
\mediac{\left( \bm{\nu},\bm{u}^\mathrm{sp}\right)^{k}} 
\end{eqnarray} 
\begin{eqnarray}
\label{CSP}
C_{N}= 
\frac{ R(\bm{\nu}^\mathrm{sp}) }
{ \sqrt{ \left| 
\det \nabla^2 \phi(\bm{\nu}^\mathrm{sp},\bm{q}^\mathrm{sp})
\right| }}, 
\end{eqnarray}
$\nabla^2 \phi(\bm{\nu},\bm{q})$ being the Jacobian of $\phi$ 
with elements
\begin{eqnarray*}
&&\partial_{q_\alpha} \partial_{q_\beta} \phi(\bm{\nu},\bm{q}) =
-\sum_{k=2}^{\infty} \frac{N^{k-1}}{(k-2)!}
\mediac{\left( \bm{\nu},i\bm{q}\right)^{k-2}\nu_{\alpha}\nu_{\beta}}
\\
&&\partial_{\nu_\alpha} \partial_{\nu_\beta} \phi(\bm{\nu},\bm{q}) =
-\frac{v_{\alpha}v_{\beta}}{(\bm{\nu},\bm{w})} 
\\
&&\partial_{q_\alpha} \partial_{\nu_\beta} \phi(\bm{\nu},\bm{q}) =
\partial_{\nu_\alpha} \partial_{q_\beta} \phi(\bm{\nu},\bm{q}) =
-i\delta_{\alpha,\beta},
\end{eqnarray*} 
for $\alpha,\beta \in \mathscr{H}$.
It is convenient to introduce two $m \times m$ matrices $\bm{\Sigma}$
and $\bm{A}$ with elements 
\begin{eqnarray}
\label{SIGMA}
\Sigma_{\alpha,\beta} &=&  
-\partial_{q_\alpha} \partial_{q_\beta} 
\phi(\bm{\nu}^\mathrm{sp},\bm{q}^\mathrm{sp}) 
\\
A_{\alpha,\beta} &=& 
-\partial_{\nu_\alpha} \partial_{\nu_\beta} 
\phi(\bm{\nu}^\mathrm{sp},\bm{q}^\mathrm{sp}) .
\label{AMAT}
\end{eqnarray} 
In terms of these matrices Eq.~(\ref{CSP}) becomes
\begin{eqnarray}
\label{CSP1}
C_{N}=\frac{1}{\sqrt{\left| \det (\bm{1} + \bm{\Sigma}\bm{A} ) \right| }}
\frac{1}{\sqrt{2 \pi N \epsilon^2 (\bm{\nu}^\mathrm{sp},\bm{w}^\mathrm{sp})}}.
\end{eqnarray}
Note that, in the case $\hat{V}\equiv 0$, the matrix $\bm{A}$ is uniform
and, due to the summation rules (\ref{c.c.f.1}) and (\ref{c.c.f.}), 
$\bm{\Sigma}\bm{A} = \bm{0}$.
In general, the same summation rules imply
$\det (\bm{\Sigma}\bm{A})=0$, so that
$\det (\bm{1} + \bm{\Sigma}\bm{A})  \simeq 1$ up to 
second order terms in $\bm{A}$.

\subsection{Resumming the canonical series}
\label{resumming}
In this section, we evaluate the expectation 
$\E \left( {\cal M}^{t}_{\bm{n}_0} \right)$ by resumming 
the series of Eq.~(\ref{EXPANSION}).
We will substitute the series over $N$ with the integral
\begin{eqnarray}
\label{PSIINT}
\E \left( \mathcal{M}^t_{\bm{n}_0} \right) = 
\int dN 
C_{N} \exp\left[ \psi(N) \right],
\end{eqnarray} 
and we will perform the integration again with a saddle-point technique. 
At the end, we will demonstrate the asymptotic exactness, 
in the limit $t\to \infty$, 
of the saddle-point integration, 
as well as of the substitution of the series with the corresponding integral,  
by showing that in this limit Eq.~(\ref{PSIINT}) 
takes exponentially leading contributions
from values of $N$ in the range 
$[\epsilon t -\sqrt{\epsilon t}, \epsilon t+\sqrt{\epsilon t}]$. 

Before evaluating the stationarity condition for $\psi(N)$,
an important comment is in order. 
For any finite value of $N$, let us introduce the rescaled cumulants 
of order $k$, in a compact notation $\bm{\Sigma}^{(N;k)}$, 
which is defined as the tensor of rank $k$ with components 
\begin{eqnarray}
\label{RCUMN}
\Sigma^{(N;k)}_{\alpha_1,\ldots,\alpha_k} =
N^{k-1} \mediac{\nu_{\alpha_1} \ldots \nu_{\alpha_k}},
\end{eqnarray}
$\alpha_1, \ldots, \alpha_k \in \mathscr{H}$.
Let $\bm{\Sigma}^{(k)}$ be the tensor of the asymptotic values 
of the rescaled cumulants in the limit $N\to \infty$ 
\begin{eqnarray}\fl
\label{RCUM}
\Sigma^{(k)}_{\alpha_1,\ldots,\alpha_k} =
\lim_{N\to \infty}
N^{k-1} \mediac{\nu_{\alpha_1} \ldots \nu_{\alpha_k}}=
\lim_{N\to \infty}
\frac{1}{N} \mediac{N_{\alpha_1} \ldots N_{\alpha_k}}.
\end{eqnarray}
These limits exist and are finite since  
the irreducible and finite Markov chain formed by the evolving configurations 
has a finite correlation length $N_c$ with respect to the number of jumps
(see Eq. (\ref{CUMULANTSDECAY})).
Up to corrections exponentially small in $N/N_c$,
for $N \gg N_c$ we can use the effective rule
\begin{eqnarray}
\label{RULE}
\partial_N \Sigma^{(N;k)}_{\alpha_1,\ldots,\alpha_k} =0.
\end{eqnarray}

By using the rule (\ref{RULE}), 
the derivative of $\psi(N)$ with respect to $N$ is
\begin{eqnarray*}
\label{DPSI}
\partial_N \psi(N) &=&  
\sum_{k=1}^{\infty} \frac{N^{k-1}}{k!}
\mediac{\left( \bm{\nu},\bm{u}^\mathrm{sp}\right)^{k}}  
\\ &&
+ \partial_N x_{0}^{\mathrm{sp}} 
\Bigg[ t - N
\sum_{k=1}^{\infty} \frac{N^{k-1}}{(k-1)!}
\mediac{\left( \bm{\nu},\bm{u}^\mathrm{sp}\right)^{k-1}
(\bm{\nu},\bm{v}^\mathrm{sp})}
\Bigg].  
\end{eqnarray*}
On the other hand, from Eq.~(\ref{SPEQ}) we have 
\begin{eqnarray*}
\label{DPSI1}
\sum_{k=1}^{\infty} \frac{N^{k-1}}{(k-1)!}
\mediac{\left( \bm{\nu},\bm{u}^\mathrm{sp}\right)^{k-1}
(\bm{\nu},\bm{v}^\mathrm{sp})}
= (\bm{\nu}^\mathrm{sp},\bm{v}^\mathrm{sp}),  
\end{eqnarray*}
so that, by using $N(\bm{\nu}^\mathrm{sp},\bm{v}^\mathrm{sp})=t$,
we find
\begin{eqnarray}
\label{DPSI2}
&& \partial_N \psi(N) =  
\sum_{k=1}^{\infty} \frac{N^{k-1}}{k!}
\mediac{\left( \bm{\nu},\bm{u}^\mathrm{sp}\right)^{k}}.  
\end{eqnarray}
In conclusion, the stationarity condition, $\partial_N \psi(N^\mathrm{sp})=0$, 
gives
\begin{eqnarray}
\label{E1}
\left. 
\sum_{k=1}^{\infty} \frac{N^{k-1}}{k!}
\mediac{\left( \bm{\nu},\bm{u}^\mathrm{sp}\right)^{k}} 
\right|_{N=N^\mathrm{sp}} \hspace{-20pt} =0,
\end{eqnarray}
which in terms of the rescaled cumulants (\ref{RCUMN}) reads
\begin{eqnarray}
\label{E2}
\left. \sum_{k=1}^{\infty}\frac{1}{k!}
\sum_{\alpha_1 \in \mathscr{H}}
\ldots
\sum_{\alpha_k \in \mathscr{H}}
\Sigma_{\alpha_1,\dots,\alpha_k}^{(N;k)}
u_{\alpha_1}^{\mathrm{sp}} \ldots u_{\alpha_k}^{\mathrm{sp}}
\right|_{N=N^\mathrm{sp}} \hspace{-20pt} =0.
\end{eqnarray}
Equation (\ref{E2}) determines the saddle point $N^\mathrm{sp}$
for the chosen value of the time $t$. 

From Eq.~(\ref{DPSI2}) and using the rule (\ref{RULE}), 
for the second derivative of $\psi(N)$ we find
\begin{eqnarray}
\label{d2psidN}
\partial^2_N \psi (N) = 
- ( \bm{\nu}^\mathrm{sp}, \bm{v}^\mathrm{sp})
\partial_N x_{0}^{\mathrm{sp}},
\end{eqnarray}
where 
\begin{eqnarray}
\label{dx0dN}
\partial_N x_{0}^{\mathrm{sp}} =
\frac{1}{N}  
\frac{( \bm{\nu}^\mathrm{sp}, \bm{v}^\mathrm{sp}) }
{
(\bm{\nu}^\mathrm{sp}, \bm{w}^\mathrm{sp}) +
(\bm{\Sigma}\bm{v}^\mathrm{sp},\bm{v}^\mathrm{sp})
}.\quad
\end{eqnarray}
Equation (\ref{dx0dN}) has been obtained by evaluating the total derivate 
of the saddle-point equation 
$N(\bm{\nu}^\mathrm{sp},\bm{v}^\mathrm{sp})=t$ with respect to $N$,
\begin{eqnarray*}
( \bm{\nu}^\mathrm{sp}, \bm{v}^\mathrm{sp}) +
N \left[
( \partial_{x_{0}^{\mathrm{sp}}} \bm{\nu}^\mathrm{sp}, \bm{v}^\mathrm{sp}) -
( \bm{\nu}^\mathrm{sp}, \bm{w}^\mathrm{sp})
\right]
\partial_N x_{0}^{\mathrm{sp}} =0,
\end{eqnarray*}
and by observing that, according to Eq.~(\ref{SPEQ}), we have
\begin{eqnarray*}
( \partial_{x_{0}^{\mathrm{sp}}} \bm{\nu}^\mathrm{sp} ,\bm{v}^\mathrm{sp} )
\! &=& \!
- \sum_{k=2}^{\infty} \frac{N^{k-1}}{(k-2)!}
\mediac{\left( \bm{\nu},\bm{u}^\mathrm{sp}\right)^{k-2}
(\bm{\nu},\bm{v}^\mathrm{sp})^2}
\nonumber \\ &=&
- (\bm{\Sigma}\bm{v}^\mathrm{sp},\bm{v}^\mathrm{sp}) .
\end{eqnarray*}

To evaluate the integral (\ref{PSIINT}), 
we approximate $\psi(N) \simeq \psi(N^\mathrm{sp}) + \frac12 
\partial^2_N \psi (N^\mathrm{sp}) (N-N^\mathrm{sp})^2$
and $C_N \simeq C_{N^\mathrm{sp}}$.
The remaining Gaussian integration gives the following result
\begin{eqnarray}
\label{EMt}
\E \left( \mathcal{M}^t_{\bm{n}_0} \right) = 
\left. \sqrt{
\frac{ 1 + \mathrm{tr} (\bm{\Sigma}\bm{A} )} 
{\left| \det (\bm{1} + \bm{\Sigma}\bm{A} ) \right| }}
~ \frac{e^{x_{0}^{\mathrm{sp}} t}}
{ \epsilon (\bm{\nu}^\mathrm{sp}, \bm{v}^\mathrm{sp})}
\right|_{N=N^{\mathrm{sp}}}.
\end{eqnarray} 
The matrices $\bm{\Sigma}$ and $\bm{A}$ are defined by Eqs. 
(\ref{SIGMA}) and (\ref{AMAT}) and their components explicitly read
\begin{eqnarray}
\Sigma^{}_{\alpha,\beta} &=& \Sigma^{(N;2)}_{\alpha,\beta} 
+ 
\sum_{k=1}^{\infty} \frac{1}{k!} 
\sum_{\alpha_1 \in \mathscr{H}}
\ldots
\sum_{\alpha_k \in \mathscr{H}}
\Sigma_{\alpha,\beta,\alpha_1,\dots,\alpha_k}^{(N;k+2)}
u_{\alpha_1}^{\mathrm{sp}} \ldots u_{\alpha_k}^{\mathrm{sp}} , \qquad
\\
A_{\alpha,\beta} &=& 
\frac{v_{\alpha}^{\mathrm{sp}}v_{\beta}^{\mathrm{sp}}}
{\left( \bm{\nu}^{\mathrm{sp}},\bm{w}^{\mathrm{sp}} \right)} . 
\end{eqnarray}

Equation (\ref{EMt}) represents our final expression for the 
matrix elements (\ref{TheFormulab}).
It is valid for $t$ large but finite with $N^{\mathrm{sp}}(t)$ determined
by Eq.~(\ref{E2}).  
It is simple to show that in the limit $t \to \infty$, 
Eq.~(\ref{E2}) may admit a solution only if $N^{\mathrm{sp}}(t)$ 
grows proportionally to $t$.
In fact, from Eq.~(\ref{WSADDLE2}) for arbitrary $t$ and $N$ we have that  
\begin{eqnarray}
\label{bds}
\nu^{\mathrm{sp}}_{V_\mathrm{min}} \frac{N+1}{t} 
< x_0^{\mathrm{sp}} + V_\mathrm{min} \leq\frac{N+1}{t},
\end{eqnarray}
where $V_\mathrm{min}$ is the smallest value of $V\in\mathscr{V}$ such that 
$\nu^{\mathrm{sp}}_{V_\mathrm{min}}>0$. 
Note that this value of $V$ exists due to the properties 
$0 \leq\nu^{\mathrm{sp}}_{\alpha}\leq 1$ for any
$\alpha\in \mathscr{V}$ and
$\sum_{\alpha\in \mathscr{V}} \nu^{\mathrm{sp}}_{\alpha}=1$.
Therefore, for $N=N_\mathrm{sp}(t)$
all the components $u_V=-\log[(x_0^{\mathrm{sp}}+V)/\varepsilon]$ 
diverge when $t\to \infty$ if 
$\lim_{t\to \infty} N^{\mathrm{sp}}(t)/t = \infty$, 
whereas if $\lim_{t\to \infty} N^{\mathrm{sp}}(t)/t = 0$ 
the only diverging component
is $u^{}_{V_\mathrm{min}}$.
In both cases, Eq.~(\ref{E2}) does not have solution for $t\to \infty $.
In this limit, therefore, we must have $N^{\mathrm{sp}}(t) \sim t$. 
This condition, which from a physical point of view simply expresses
the proportionality between the time and the length of the trajectories 
most contributing to the expectation,
implies two important consequences.
First, since the peak of the Gaussian at $N=N^\mathrm{sp}(t)$ 
moves to infinity linearly with $t$,  
whereas its width increases only as $\sqrt{t}$,
see Eqs.~(\ref{d2psidN}) and (\ref{dx0dN}),
the result (\ref{EMt}) becomes asymptotically exact for $t\to \infty$.
Second, in Eq.~(\ref{E2}) we can substitute $\bm{\Sigma}^{(N;k)}$ with their
asymptotic values $\bm{\Sigma}^{(k)}$. 
In this way, Eq.~(\ref{E2}) can be read as a time-independent equation
that determines the quantity 
\begin{eqnarray}
\label{E0B}
E_{0B} = - \lim_{t \to \infty}
 \left. x_{0}^{\mathrm{sp}} \right|_{N=N^\mathrm{sp}(t)} 
\end{eqnarray}
as a function of the asymptotic rescaled cumulants
$\bm{\Sigma}^{(k)}$, $k\geq 1$.
By comparing Eqs.~(\ref{E0}) and (\ref{EMt}), we realize
that this quantity is the ground-state energy of the considered system.
In conclusion, for systems of hard-core bosons 
the ground-state energy $E_{0B}$ is determined by the scalar equation
\begin{eqnarray}
\label{E0BEQ}
\sum_{k=1}^{\infty}
\frac{1}{k!}
\sum_{\alpha_1\in\mathscr{H}}
\ldots
\sum_{\alpha_k\in\mathscr{H}}
\Sigma_{\alpha_1,\ldots,\alpha_k}^{(k)}
u_{\alpha_1}(E_{0B}) \dots u_{\alpha_k}(E_{0B}) =0 ,
\end{eqnarray}
where
\begin{eqnarray}
\bm{u}^\mathrm{T} (E_{0B})
= (\ldots -\log[(-E_{0B}+V)/\epsilon] \ldots;
\ldots \log T \ldots).
\end{eqnarray}
If we truncate the series in Eq.~(\ref{E0BEQ}) to the second order, 
we recover the Gaussian approximation of Ref.~\cite{OP1} 
(there $\bm{\Sigma}^{(1)}$ and $\bm{\Sigma}^{(2)}$ were indicated as
$\overline{\bm{\nu}}$ and $\bm{\Sigma}$, respectively).
Furthermore, as in \cite{OP1} 
the $\hat{V}\equiv 0$ case can be solved explicitly
and we get the following exact solution for the ground-state energy, 
$E_{0B}^{(0)}$, of a non interacting hard-core boson system
\begin{eqnarray}
E_{0B}^{(0)} = -\epsilon \exp\left[\sum_{k=1}^{\infty}\frac{1}{k!}
\sum_{T_1 \in\mathscr{T}}
\ldots 
\sum_{T_k \in\mathscr{T}}
\Sigma^{(k)}_{T_1,\ldots,T_k}
\log T_1 \dots \log T_k
\right].
\end{eqnarray}
For a generic $\hat{V}$,
 Eq.~(\ref{E0BEQ}), with the series truncated at an
arbitrary finite order $k_\mathrm{max}$, 
can be solved numerically by the bisection method
using the bounds  $E_{0B}^{(0)} \leq  E_{0B} \leq 0$.

Equation (\ref{E0BEQ}) allows,
via the Hellman-Feynman theorem (\ref{HF}), 
also to evaluate the expectation of any other operator 
in the ground state of the chosen Hamiltonian
(see Ref.~\cite{OP1} for more details).
Moreover, it is clear that the cumulants $\bm{\Sigma}^{(k)}$ 
depend only on the structure of the system Hamiltonian, 
not on the values of the Hamiltonian parameters.
Therefore, once the $\bm{\Sigma}^{(k)}$ are known,  
all the evaluated ground-state expectations are analytical 
functions of the Hamiltonian parameters.

\section{Numerical evaluation of the cumulants and example cases}
\label{nresults}
In this section, we discuss the numerical evaluation of the cumulants.
We also apply our method 
to some example cases and compare the ground-state energies
obtained by Eq.~(\ref{E0BEQ}) with those from
exact numerical calculations.

In our approach, the starting point is the evaluation of the
asymptotic values $\bm{\Sigma}^{(k)}$ 
of the rescaled cumulants $\bm{\Sigma}^{(N;k)}$.
According to Eq.~(\ref{c.c.f.R}), the latter are obtained in terms of the 
statistical moments $\media{N_{\alpha_1} \ldots N_{\alpha_k}}$, 
which are easily sampled by generating $M$ random 
trajectories branching from the initial configuration $\bm{n}_0$, 
\textit{i.e.}   
\begin{eqnarray}
\label{sample}
\media{N_{\alpha_1} \ldots N_{\alpha_k}} &=&
\sum_{\bm{\mu}} \mathcal{P}_N(\bm{\mu}) N_{\alpha_1} \ldots N_{\alpha_k} 
\nonumber \\ &\simeq&
\frac{1}{M} \sum_{p=1}^M 
N_{\alpha_1}^{(p)} \ldots N_{\alpha_k}^{(p)},
\end{eqnarray}
where $N_{\alpha}^{(p)}$ is the multiplicity $N_\alpha$, 
$\alpha \in \mathscr{H}$, of the $p$th trajectory.
The number $M$ of trajectories must be chosen 
larger than a critical value $M_\varepsilon(N,k)$,
which depends on the statistical precision $\varepsilon$ required
in the evaluation of the rescaled cumulants $\bm{\Sigma}^{(N;k)}$.

\begin{figure}[t]
\centering
\psfrag{2x3 3+3}[l][][0.8]{$2 \times 3$ $N_p=3$}
\psfrag{4x4 8+8}[l][][0.8]{$4 \times 4$ $N_p=8$}
\psfrag{xx}[t][]{$h$}
\psfrag{yy}[b][]{$\left| C_{V=\gamma,V=\gamma}(h) \right|$}
\includegraphics[width=0.7\columnwidth,clip]{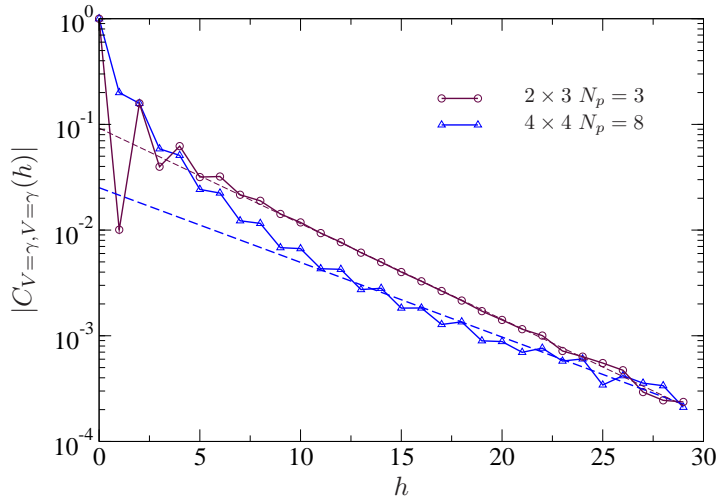}
\caption{
Absolute value of the normalized correlation function
$C_{V=\gamma,V=\gamma}(h)$
as a function of the jump interval $h=h_2-h_1$ 
for two different FNU hard-core boson Hubbard models. 
The straight dashed lines are the results of a fit 
in the range $10 \leq h \leq 29$ with a function of the form 
${\rm const} \times \exp(-h/N_c)$.
We have $N_c=4.80$ and $N_c=6.14$ in the
$2 \times 3$ $N_p=3$ and 
$4 \times 4$ $N_p=8$ cases, respectively.
Each correlation function has been evaluated by using $M=0.5 \times 10^9$
independent trajectories.}
\label{cumulants.correl.eps}
\end{figure}

The length $N$ of the $M$ trajectories is chosen sufficiently large 
for the asymptotic behavior of $\bm{\Sigma}^{(N;k)}$ to be established.
This may represent a problem since the fluctuations in Eq.~(\ref{sample}) 
grow as $N^k$ so that $M_\varepsilon(N,k)$ becomes huge for $k>1$. 
However, the evolving trajectories form a Markov chain having, see later, 
a finite correlation length $N_c$.
Therefore, $\bm{\Sigma}^{(N;k)}$ converges exponentially to 
$\bm{\Sigma}^{(k)}$ with a characteristic length of the order of $(k+1) N_c$,
$k N_c$ if the initial configuration $\bm{n}_0$ is taken randomly
distributed according to the invariant measure of the Markov chain.   
As shown in the example case of Fig.~\ref{cumulants.correl.eps}, 
for a large class of models we have observed that the correlation length
$N_c$ exists and grows no more than linearly with the size of the system.
The evaluation of the cumulants is thus feasible even for large systems. 

A possible numerical limitation in the evaluation of the cumulants 
of large order is collecting and updating all the 
components of $\bm{\Sigma}^{(k)}$, 
whose number is $m^k$ with $m$, the cardinality of $\mathscr{H}$, 
that grows with the size of the system.
Basically, this kind of difficulty is related to the CPU capability 
of simultaneously updating many addresses of memory, not to the CPU speed,
and can be reduced by vectorization tools.
The difficulty can be lowered also exploiting the invariance of 
the components $\Sigma^{(k)}_{\alpha_1,\ldots,\alpha_k}$ under 
permutation of any pair of the $\alpha$ indices.   
In this way, only the components
$\Sigma^{(k)}_{\alpha_1,\ldots,\alpha_k}$ with 
$\alpha_1 \leq \alpha_2 \leq \ldots \leq \alpha_k$ are 
sampled according to Eq.~(\ref{sample}).
This introduces an error in the summation rules (\ref{c.c.f.1}) 
and (\ref{c.c.f.}), 
which, on the other hand, are identically satisfied if all the components of 
$\bm{\Sigma}^{(k)}$ are sampled.  
However, more than a drawback this error represents an advantage, 
which allows to set the critical number $M_\varepsilon$
of sampling trajectories in a simple way.
In fact, the determination of the cumulants with a statistical precision
$\varepsilon$ implies the summation rules (\ref{c.c.f.1}) and (\ref{c.c.f.}) 
to be satisfied with the same precision.

In the following, we report on simulations performed in the case of the 
first-neighbor uniform (FNU) hard-core boson Hubbard model 
defined by the Hamiltonian
\begin{eqnarray}
\label{FNU}
\hat{H} = - \eta 
\sum_{(i,j) \in \Gamma} \sum_{\sigma=\ua\da}
\left( \hat{c}^\dag_{i\sigma} \hat{c}^{}_{j\sigma}
+\hat{c}^\dag_{j\sigma} \hat{c}^{}_{i\sigma} \right) 
+ \gamma
\sum_{i \in \Lambda}
\hat{c}^\dag_{i\ua} \hat{c}^{}_{i\ua} \hat{c}^\dag_{i\da} \hat{c}^{}_{i\da},
\end{eqnarray}
where 
$\Gamma = 
\{(i,j): \mbox{ $i<j \in \Lambda$ and $i,j$ first neighbors}\}$.
We set the reference energy $\epsilon$ to the value of 
the hopping parameter $\eta$. 
We have considered two-dimensional systems with periodic boundary conditions,
$L_x \times L_y$ sites and $N_\ua=N_\da=N_p$ particles per spin.
For this model, the set $\mathscr{V}$ has elements 
$0,\gamma,2\gamma,\ldots,N_p\gamma$,
whereas the set $\mathscr{T}$ is the collection of the possible values of the 
number of active links, \textit{e.g.}, $\mathscr{T}=\{12,16,20\}$ 
in the case of a $2 \times 3$ system with $N_p=3$
or $\mathscr{T}=\{8,10,12,14,16\}$ for the same system with $N_p=2$.

To check the correlation properties mentioned before, 
we studied the connected correlation functions of order 2 
\begin{eqnarray}
\label{cf}
C_{\alpha,\beta}(h_2-h_1) &=& 
\mediac{
\chi^{}_{\alpha}\left( \bm{n}^{}_{h_1} \right)
\chi^{}_{\beta}\left( \bm{n}^{}_{h_2} \right) },
\end{eqnarray}
where $h_2 \geq h_1$ and 
$\chi^{}_{\alpha}\left( \bm{n} \right)$ is the characteristic function 
taking the value 1, if $V(\bm{n})=\alpha$ and $\alpha \in \mathscr{V}$, 
or $T(\bm{n})=\alpha$ and $\alpha \in \mathscr{T}$,
and 0 otherwise.
The averages $\mediac{\cdot}$ can be sampled as indicated in 
Eq.~(\ref{sample}) by generating $M$ independent trajectories
with configurations $\bm{n}_h^{(p)}$, 
$h=0,1,\ldots,N$ and $p=1,\ldots,M$.
For $N$ sufficiently large, the correlation functions (\ref{cf})
do no longer depend on $h_1$ and $h_2$ but only on the jump interval 
$h=h_2-h_1$.
In Fig.~\ref{cumulants.correl.eps} we show the behavior of the correlation
function (normalized to 1 at $h=0$) obtained by choosing 
$\alpha$ and $\beta$ equal to the potential value $V=\gamma$
for two different FNU hard-core boson Hubbard models. 
After an initial transient, $C_{V=\gamma,V=\gamma}(h)$ 
decreases as $\exp(-h/N_c)$. 
The measurement of the correlation length by a fitting procedure 
shows that $N_c$ increases slowly with the size of the system. 
Similar results are obtained for different choices of $\alpha$ and $\beta$.

\begin{figure}[t]
\centering
\psfrag{2x3 3+3}[][][0.8]{$2 \times 3$ $N_p=3$}
\psfrag{2x3 2+2}[][][0.8]{$2 \times 3$ $N_p=2$}
\psfrag{m1}[l][][0.8]{$1$}
\psfrag{m2}[l][][0.8]{$2$}
\psfrag{m3}[l][][0.8]{$3$}
\psfrag{m4}[l][][0.8]{$4$}
\psfrag{xx}[t][]{$\gamma/\eta$}
\psfrag{yy}[b][]{$E_{0B}/(\eta N_p)$}
\includegraphics[width=0.7\columnwidth,clip]{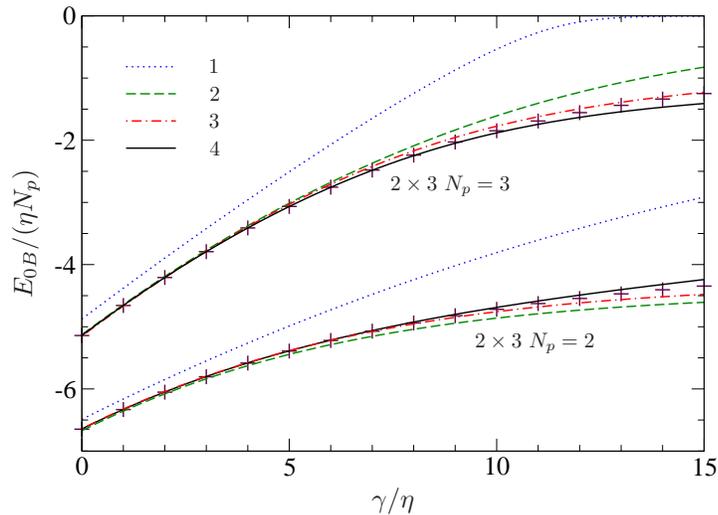}
\caption{
Ground-state energy per particle for the $2\times 3$ FNU hard-core 
boson Hubbard model \textit{versus} the interaction strength $\gamma$
with $N_p=2$ and $N_p=3$ particles per spin. 
We compare the results obtained by solving Eq.~(\ref{E0BEQ}) at
truncation orders $k_\mathrm{max}=1,2,3,4$ (different lines) 
with those from exact numerical diagonalizations ($\bm+$). 
The statistical errors associated to the cumulants, 
evaluated with $N=200$ and $M=10^7$, are negligible in this scale.} 
\label{cumulants.2x3.eps}
\end{figure}

\begin{figure}[t]
\centering
\psfrag{4x4 8+8}[][][0.8]{$4 \times 4$ $N_p=8$}
\psfrag{4x4 5+5}[][][0.8]{$4 \times 4$ $N_p=5$}
\psfrag{m1}[l][][0.8]{$1$}
\psfrag{m2}[l][][0.8]{$2$}
\psfrag{m3}[l][][0.8]{$3$}
\psfrag{m4}[l][][0.8]{$4$}
\psfrag{xx}[t][]{$\gamma/\eta$}
\psfrag{yy}[b][]{$E_{0B}/(\eta N_p)$}
\includegraphics[width=0.7\columnwidth,clip]{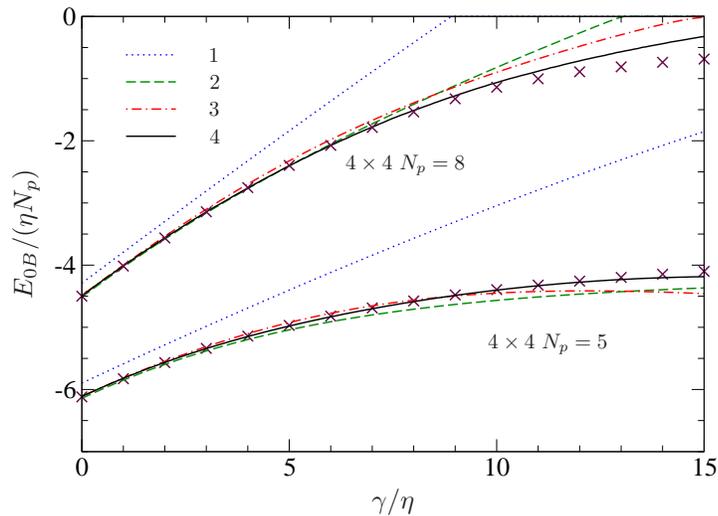}
\caption{As in Fig.~\ref{cumulants.2x3.eps} for the 
$4\times 4$ lattice with $N_p=5$ and $N_p=8$ particles per spin.
We compare the results obtained by solving Eq.~(\ref{E0BEQ}) at
truncation orders $k_\mathrm{max}=1,2,3,4$ (different lines) 
with those from quantum Monte Carlo 
simulations ($\bm\times$).
The statistical errors associated to the cumulants, 
evaluated with $N=200$ and $M=10^7$, 
and to the quantum Monte Carlo results are negligible in this scale.}
\label{cumulants.4x4.eps}
\end{figure}

In Fig.~\ref{cumulants.2x3.eps} we show the behavior of $E_{0B}$
as a function of the interaction strength $\gamma$ in a $2\times 3$ lattice
with $N_p=2$ and $N_p=3$, whereas in Fig.~\ref{cumulants.4x4.eps} 
we consider a $4\times 4$ lattice with $N_p=5$ and $N_p=8$.
In these figures, we compare the energies obtained from Eq.~(\ref{E0BEQ}) 
by truncating the cumulant expansion at the order $k_\mathrm{max}=1,2,3,4$ 
with the results from exact numerical diagonalizations 
(Fig.~\ref{cumulants.2x3.eps}) and
quantum Monte Carlo simulations (Fig.~\ref{cumulants.4x4.eps}).
For $k_\mathrm{max}=2$, we recover the results of Ref.~\cite{OP1}.
As expected, we obtain better and better agreement with the exact energies 
as the truncation order $k_\mathrm{max}$ is increased.

The number of cumulants needed to obtain a given
approximation grows as the the interaction strength $\gamma$ or
the lattice size $|\Lambda|$ are increased.
As explained in the previous Sections and anticipated in \cite{OP1},
this behavior is due to the form of the function $f(\bm{\mu})$ to be 
averaged in Eq.~(\ref{AVERAGE1}). 
In fact, $f(\bm{\mu})$ involves multiplicities 
$\bm{\mu}$ coupled with the potential values and with 
the number of active links,
which, in turn, are related to $\gamma$ and $|\Lambda|$.  
For increasing values of $\gamma$ and $|\Lambda|$, 
$f(\bm{\mu})$ becomes more and more
sensible to the large deviations of the probability density
$\mathcal{P}_N(\bm{\mu})$ and cumulants of higher and higher order must
be kept in the calculation.

\begin{figure}[t]
\centering
\psfrag{2x3 3+3}[][][0.8]{$2 \times 3$ $N_p=3$}
\psfrag{4x4 8+8}[][][0.8]{$4 \times 4$ $N_p=8$}
\psfrag{m1}[l][][0.8]{$1$}
\psfrag{m2}[l][][0.8]{$2$}
\psfrag{m3}[l][][0.8]{$3$}
\psfrag{m4}[l][][0.8]{$4$}
\psfrag{m5}[l][][0.8]{$5$}
\psfrag{xx}[t][]{$\gamma/\eta$}
\psfrag{yy}[b][]{$E_{0B}/(\eta N_p)$}
\includegraphics[width=0.7\columnwidth,clip]{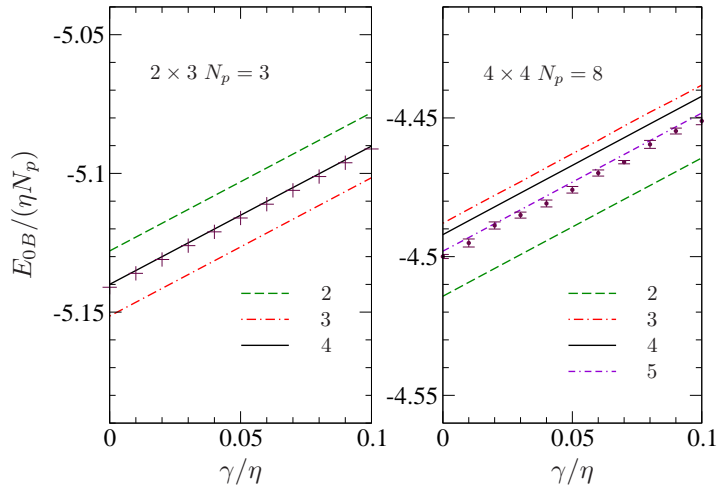}
\caption{Enlargement of Fig. \ref{cumulants.2x3.eps} (left panel) 
and Fig. \ref{cumulants.4x4.eps} (right panel) in the small $\gamma$ 
region of the systems at half filling considered there.  
We show the results obtained by solving Eq.~(\ref{E0BEQ})
at truncation orders $k_\mathrm{max}=2,3,4$ (different lines, left panel)
and $k_\mathrm{max}=2,3,4,5$ (different lines, right panel)
in comparison with those from exact numerical diagonalizations 
($\bm+$, left panel) and quantum Monte Carlo simulations 
(dots with error bars, right panel).
The statistical errors associated to the cumulants are negligible.}
\label{cumulants.enlarg.eps}
\end{figure}

In Fig.~\ref{cumulants.enlarg.eps},
which is an enlargement of Fig.~\ref{cumulants.2x3.eps} (left panel)
and Fig.~\ref{cumulants.4x4.eps} (right panel) in the small
$\gamma$ region of the systems at half filling considered there,
we can better appreciate the convergence of the the solutions 
of Eq.~(\ref{E0BEQ}), for increasing values of $k_\mathrm{max}$, 
toward the exact energies. 
In all Figs.~\ref{cumulants.2x3.eps}-\ref{cumulants.enlarg.eps}, 
the statistical errors associated to the measurement of the cumulants 
are negligible in the scales considered.

\section{Conclusions}
\label{conclusions}
By using saddle-point techniques and a cumulant expansion theorem, 
we have exploited an exact probabilistic representation of the
quantum dynamics in a lattice to evaluate 
the matrix elements of the evolution operator of a system
of hard-core bosons in the limit of long times. 
The approach yields a simple scalar equation for the ground-state energy
in terms of the asymptotic cumulants $\bm{\Sigma}^{(k)}$ of the values of 
the potentials, $V$, and of the kinetic quantities, $T$, 
assumed by the system during its long-time evolution.  
Since the cumulants depend only on the structure of the system
Hamiltonian, once they are known, 
this equation provides the ground-state energy 
and, via the Hellman-Feynman theorem (\ref{HF}),
the ground-state expectation of any other operator, 
analytically as a function of the Hamiltonian parameters.
In contrast, quantum Monte Carlo methods require, 
due to the unavoidable branching or reconfiguration techniques
(see \cite{OP3} and \cite{CALANDRASORELLA} and references therein),
different simulations for different values of the parameters.

The analytical character of the present approach suggests 
many potential applications. 
Here, we briefly envisage some of them. 

\textit{i)} The knowledge of the ground-state energy as a function
of the Hamiltonian parameters, $E_{0B}=E_{0B}(\bm{\xi})$, allows
in principle the determination of ground-state quenched-averages,
$\int f(E_{0B}(\bm{\xi}))dP_{\bm{\xi}}$, 
of crucial interest in the study of disordered systems. 
Here, $P_{\bm{\xi}}$ is a given probability density of the disorder 
parameters $\bm{\xi}$. 
Note that when the dimension of the space of the Hamiltonian parameters 
is larger than 1, the evaluation of the above integral 
by a quantum Monte Carlo approach may be a hard, if not unmanageable,
numerical task.

\textit{ii)} Even though we have specialized the study 
to the easiest case, 
namely that of hard-core bosons in the absence of a magnetic field, 
as evidenced at the beginning of Section III and in Section III A, 
our approach is general and not limited to hard-core boson systems. 
Fermions and bosons in a magnetic field may in principle be treated 
in a similar way, provided that we properly take into account also the 
multiplicities $N_{\lambda}$, see Eq.~(\ref{MULTIPLICITY-L}).
This possibility is of great interest since, as well known,
fermions and bosons in a magnetic field are both affected 
by the so called sign (or phase) problem, 
which in practice inhibits the accomplishment of 
unbiased quantum Monte Carlo simulations. 	

\textit{iii)} We note that our approach is also analytical  
in the time parameter, $t$.
By properly taking into account the derivatives of the cumulants
with respect to the number of jumps, $\partial_N \bm{\Sigma}^{(N;k)}$, 
one can obtain not only the asymptotic behavior of the matrix elements, 
Eq.~(\ref{EMt}), but also their behavior at finite times, $t$, 
both imaginary or real. 
The latter possibility constitutes another chance of great interest 
because in general, due to the presence of oscillating terms,
also the real time behavior is affected by a sort of sign problem
and the quantum Monte Carlo simulations are reliable only for
short times \cite{OP3}.
Of course, the complete knowledge of the time behavior would imply 
that of the excited states.

\textit{iv)} Finally, we mention a different possible application
of the result (\ref{E0BEQ}). 
This equation can be exploited in a Monte Carlo framework for
effectively sampling the ground-state energy (work in progress). 
Essentially, the better efficiency of this Monte Carlo method,
compared with that directly deduced from Eq.~(\ref{TheFormulaa}), 
see Ref.~\cite{OP3} for details, 
follows from the fact that the stochastic times of the 
original probabilistic representation have been 
analytically integrated out, as done in Section III B, 
so that the fluctuations are necessarily reduced.

\section*{Acknowledgments}
We thank F. Cesi and G. Jona-Lasinio for enlightening discussions 
and a critical reading of the manuscript.
This work was supported in part by Cofinanziamento MIUR 
protocollo 2002027798$\_$001.

\appendix
\section*{Appendix A. Proof of the cumulant summation rules}
\label{csr-proof}
\addcontentsline{toc}{section}{Appendix A. Proof of the cumulant summation rules}
\setcounter{section}{1}
In this Appendix we prove the relations (\ref{c.c.f.1}) and (\ref{c.c.f.}).
Due to the constraints (\ref{CONSTRAINTS}),
it is trivial to see that 
the statistical moments (non-connected correlation functions) of order $k$, 
$\media{\nu_{\alpha_1}\dots\nu_{\alpha_k}}$,
$ \alpha_1,\dots, \alpha_{k} \in\mathscr{H}$,
satisfy the following summation rules
\begin{eqnarray}
\label{MOMENTS1}
\sum_{\alpha \in \mathscr{V}}\media{\nu_\alpha}=
\sum_{\alpha \in \mathscr{T}}\media{\nu_\alpha}=1,
\end{eqnarray}
for $k=1$, whereas for $k>1$
\begin{eqnarray}
\label{MOMENTS}
\sum_{\alpha_k \in \mathscr{V}}
\media{\nu_{\alpha_1}\dots\nu_{\alpha_k}} &=&
\sum_{\alpha_k \in \mathscr{T}}
\media{\nu_{\alpha_1}\dots\nu_{\alpha_k}} 
\nonumber\\ &=& 
\media{\nu_{\alpha_1}\dots\nu_{\alpha_k-1}}.
\end{eqnarray}
Since $\mediac{\nu_\alpha} = \media{\nu_\alpha}$, 
Eq. (\ref{MOMENTS1}) coincides with Eq. (\ref{c.c.f.1}).

To demonstrate Eq. (\ref{c.c.f.}), let us introduce the short notation
\begin{eqnarray}
\label{I}
m( I^{(k)})=\media{\nu_{\alpha_1}\dots\nu_{\alpha_k}} \\
s(I^{(k)})=\mediac{\nu_{\alpha_1}\dots\nu_{\alpha_k}}, 
\end{eqnarray}
where $I^{(k)}=\{1,\dots,k\}$ is the set of integers that appear
as subscripts in $\alpha_1\dots\alpha_k$.
More generally, for any nonempty subset $I_p$ of $I^{(k)}$,
$I_p=\{p_1,p_2,\dots\}$, we define $m(I_p)$ and $s(I_p)$ as
\begin{eqnarray}
\label{Ip}
m( I_p)=\media{\nu_{\alpha_{p^{}_1}}\nu_{\alpha_{p^{}_2}}\dots} \\
s(I_p)=\mediac{\nu_{\alpha_{p^{}_1}}\nu_{\alpha_{p^{}_2}}\dots}. 
\end{eqnarray}
The cumulants can be defined implicitly in terms 
of the moments according to the relation \cite{SHY}
\begin{eqnarray}
\label{c.c.f.R}
m(I^{(k)})=
\sum_{ \cup_{p} I_p=I^{(k)} }
~ \prod_{p} s(I_p),
\end{eqnarray}
where the sum is extended to all the unordered decompositions
of the set $I^{(k)}$ in disjoint nonempty sets $I_p$ 
such that $\cup_{p} I_p=I^{(k)}$.

Let us proceed inductively and suppose that Eqs.~(\ref{c.c.f.}) 
hold for any value of the order $k$ such that $2 \leq k \leq n-1$,
\textit{i.e.} 
\begin{eqnarray}
\label{c.c.f.Rh}
\sum_{\alpha_{k}}s(I^{(k)})=0, \qquad 2 \leq k \leq n-1,
\end{eqnarray}
where the sum runs either over the sets $\mathscr{V}$ or $\mathscr{T}$.

At the order $n$, we rewrite Eq.~(\ref{c.c.f.R}) as
\begin{eqnarray}\fl
\label{c.c.f.R1}
m(I^{(n)}) = 
s(I^{(n)})+
\sum_{ \cup_{p} I_p=I^{(n-1)} }
~\prod_{p} s(I_p)s(\{n\})
\nonumber \\ +
\sum_{\cup_p I_p=I^{(n)},~I_p\neq I^{(n)},~I_p\neq\{n\}}
\prod_{p} s(I_p) ,
\end{eqnarray}
where $\{n\}$ is the set having only the element $n$ and 
$s(\{n\}) =\mediac{\nu_{\alpha_n}}$.
By summing over the index $\alpha_{n}$, 
for $\alpha_{n} \in \mathscr{V}$ or $\alpha_{n} \in \mathscr{T}$,
and using the relations (\ref{MOMENTS}) together with 
$\mediac{\nu_{\alpha_n}} = \media{\nu_{\alpha_n}}$, we get
\begin{eqnarray}\fl
\label{c.c.f.R2}
m(I^{(n-1)}) =
\sum_{\alpha_{n}}s(I^{(n)})+
\sum_{ \cup_{p} I_p=I^{(n-1)} }
~\prod_{p} s(I_p) 
\nonumber \\ + 
\sum_{\cup_{p} I_p=I^{(n)},~I_p\neq I^{(n)},~I_p\neq\{n\}}
\prod_{p} \sum_{\alpha_{n}}s(I_p).
\end{eqnarray}
According to Eq.~(\ref{c.c.f.R}), 
the second term in the r.h.s. of Eq.~(\ref{c.c.f.R2}) 
is equal to $m(I^{(n-1)})$
whereas the third term involves only sets $I_p$ with 
$\left| I_p \right| \leq n-1$
so that by using the inductive hypothesis, Eqs.~(\ref{c.c.f.Rh}), we have
\begin{eqnarray}
\label{c.c.f.R3}
\sum_{\alpha_{n}}s(I^{(n)})=0.
\end{eqnarray}
Finally, it is easy to verify by direct inspection that 
in the case $k=2$ we have
\begin{eqnarray}
\label{c.c.f.R4}
\sum_{\alpha_{2}}s(I^{(2)})=0,
\end{eqnarray}
so that Eq.~(\ref{c.c.f.}) is proved.

\end{document}